Dated: 23 November 2019

# Anomalous and topological Hall effects in epitaxial thin films of the noncollinear antiferromagnet Mn$_3$Sn

James M. Taylor [*] [1], Anastasios Markou [2], Edouard Lesne [1], Pranava Keerthi Sivakumar [1], Chen Luo [3], Florin Radu [3], Peter Werner [1], Claudia Felser [2], Stuart S. P. Parkin [†] [1]

[1] Max Planck Institute of Microstructure Physics, Weinberg 2, 06120 Halle (Saale), Germany
[2] Max Planck Institute for Chemical Physics of Solids, Nöthnitzer Str. 40, 01187 Dresden, Germany
[3] Helmholtz-Zentrum Berlin for Materials and Energy, Albert Einstein Str. 15, 12489 Berlin, Germany

**Abstract**

Noncollinear antiferromagnets with a $D0_{19}$ (space group = 194, P6$_3$/*mmc*) hexagonal structure have garnered much attention for their potential applications in topological spintronics. Here, we report the deposition of continuous epitaxial thin films of such a material, Mn$_3$Sn, and characterize their crystal structure using a combination of x-ray diffraction and transmission electron microscopy. Growth of Mn$_3$Sn films with both (0001) *c*-axis orientation and (40$\bar{4}$3) texture is achieved. In the latter case, the thin films exhibit a small uncompensated Mn moment in the basal plane, quantified via magnetometry and x-ray magnetic circular dichroism experiments. This cannot account for the large anomalous Hall effect simultaneously observed in these films, even at room temperature, with magnitude $\sigma_{xy}$ ($\mu_0 H$ = 0 T) = 21 $\Omega^{-1}$ cm$^{-1}$ and coercive field $\mu_0 H_c$ = 1.3 T. We attribute the origin of this anomalous Hall effect to momentum-space Berry curvature arising from the symmetry-breaking inverse triangular spin structure of Mn$_3$Sn. Upon cooling through the transition to a glassy ferromagnetic state at around 50 K, a peak in the Hall resistivity close to the coercive field indicates the onset of a topological Hall effect contribution, due to the emergence of a scalar spin chirality generating a real-space Berry phase. We demonstrate that the polarity of this topological Hall effect, and hence the chiral-nature of the noncoplanar magnetic structure driving it, can be controlled using different field cooling conditions.

---

[*] james.taylor@mpi-halle.mpg.de
[†] stuart.parkin@mpi-halle.mpg.de

Antiferromagnets (AF) are of interest for spintronic applications [1], in particular topological materials [2], that can provide the large read-out signals required through electrical signatures such as the intrinsic anomalous Hall effect (AHE) [3]. $Mn_3Sn$ is a noncollinear AF with Mn moments arranged in hexagonal crystal planes [4], which exhibits an inverse triangular spin structure resulting from a combination of exchange and Dzyaloshinskii-Moriya (DM) interactions [5]. The inverse triangular AF order breaks time-reversal symmetry, thus introducing a fictitious magnetic field generated by momentum-space Berry phase, which has been theoretically predicted [6] to drive a highly anisotropic AHE [7]. This was subsequently experimentally measured in single crystals of $Mn_3Sn$ [8] (as well as in $Mn_3Ge$ [9]). Berry curvature is connected with the presence in $Mn_3Sn$ of Weyl quasiparticles [10-12], which generate an anomalous Nernst effect [13-15]. Further enhancement of the attractiveness of $Mn_3Sn$ for spintronics stems from its hosting of an intrinsic spin Hall effect [16-18] (originally discovered in the cubic noncollinear AF $Mn_3Ir$ [19]).

The inverse triangular spin texture also gives rise to a magneto-optical Kerr effect [20-22], which reveals that $Mn_3Sn$ contains AF domains possessing opposite chiralities of a cluster octupole order parameter [23] and, hence, opposite polarity of magnetotransport effects [24]. These chiral domains can be propagated [22] by coupling an external magnetic field to the small uncompensated magnetic moment [25] that is created in $Mn_3Sn$ by spins canting slightly towards magnetocrystalline easy axes [26]. This weak magnetization can freely rotate within the basal plane [5], acting to orient the entire inverse triangular spin structure into a single chiral domain state.

As $Mn_3Sn$ is cooled, its magnetic order changes depending on microstructure [27]. At 275 K, Mn-deficient samples transition to a helical magnetic phase [28]. Around 50 K, the magnetic structure changes to a 'glassy' ferromagnetic (FM) (or spin glass) state [29], regardless of composition, in which Mn moments cant out of the (0001) planes [30]. Studies in bulk samples [31], show this is accompanied by the onset of a topological Hall effect (THE) [32], attributed to a possible magnetic Skyrmion phase [33], or to chiral domain walls [34].

Recently, AHE has been measured in polycrystalline $Mn_3Sn$ films [35-37], as has a planar Hall effect in epitaxial films [38]. In this Rapid Communication, we extend these results by demonstrating both AHE and THE in epitaxial thin films of $Mn_3Sn$.

We previously grew $Mn_3Sn$ films that were epitaxial but discontinuous [39]. By changing substrate and optimizing post-annealing temperature, we succeeded in fabricating continuous $Mn_3Sn$ films (details in the Supplemental Material). Fig. 1(a) shows $2\theta$-$\theta$ x-ray diffraction (XRD) patterns measured for films deposited on $SrTiO_3$ (111) and MgO (001) substrates. In both cases, a 5 nm Ru buffer layer was used [39]. The presence of {0002} diffraction peaks indicates the growth of *c*-axis oriented films on $SrTiO_3$ (111) substrates. Lattice parameters calculated from these XRD measurements establish that these hexagonal $Mn_3Sn$ (0001) films grow fully relaxed (analysis in the Supplemental Material). Azimuthal XRD $\phi$ scans of partially in-plane (IP) peaks, shown in the inset of Fig. 1(a), demonstrate the same epitaxial relationship as in Ref. [39]: $SrTiO_3(111)[11\bar{2}][1\bar{1}0]$ || $Ru(0001)[01\bar{1}0][2\bar{1}\bar{1}0]$ || $Mn_3Sn(0001)[01\bar{1}0][2\bar{1}\bar{1}0]$, depicted schematically in Fig. 1(c).

We studied film structure at the nanoscale using transmission electron microscopy (TEM). Fig. 1(b) shows a cross-section scanning-TEM micrograph from a 70 nm film deposited on $SrTiO_3$ (111), which demonstrates explicitly the epitaxial growth of (0001) oriented $Mn_3Sn$ on a Ru (0001) buffer. The inset of Fig. 1(b) displays a wide-view scanning-TEM image of the same lamella, confirming the preparation of continuous films.

Fig. 1(a) also shows a single peak in the $2\theta$-$\theta$ XRD scan for a 60 nm film deposited on MgO (001); this corresponds to the growth of single phase hexagonal $Mn_3Sn$ with ($40\bar{4}3$) orientation. We analyzed this crystal texture by recording $\chi$-$\phi$ XRD pole figure maps around the expected positions of the {111} reflections of MgO, {0002} reflections of Ru and {0002} reflections of $Mn_3Sn$, as presented in Fig. 1(d). The {0002} reflections of Ru follow the cubic {111} MgO peaks, indicating that the hexagonal planes of Ru are tilted to lie at the same angle as the (111) planes of the substrate. The {0002} reflections of $Mn_3Sn$ in turn follow the Ru diffraction peaks, confirming that the basal planes of $Mn_3Sn$ are seeded to grow at the same angle as those in the buffer layer. This results in a structure for $Mn_3Sn$ ($40\bar{4}3$) with the [0001] crystalline direction at almost 55° to the film normal, the [2$\bar{1}$10] magnetic easy axis [26] at approximately 35° to the film normal, and the [01$\bar{1}$0] crystallographic axis lying completely IP. This epitaxial relationship is illustrated in Fig. 1(f).

The Mn$_3$Sn {0002} reflections are four-fold symmetric. This indicates four distinct crystallite orientations in which the *c*-axis, and in turn the orthogonal [01$\bar{1}$0] direction, follows one of the possible IP ⟨110⟩ MgO axes. Therefore, for a chosen macroscopic measurement direction, there will be equivalent numbers of Mn$_3$Sn crystal grains with either the [0001] or [01$\bar{1}$0] axes parallel to this.

Nevertheless, high-resolution TEM of an individual Mn$_3$Sn (40$\bar{4}$3) crystallite, as depicted in Fig. 1(e), confirms single-crystalline growth. The [01$\bar{1}$0] crystallographic axis is directed at 45° into the plane of the image. The basal planes of Mn$_3$Sn are clearly visible, whilst the [2$\bar{1}$10] magnetic easy axis is aligned almost OP. That these crystal grains coherently coalesce to form a continuous thin film is ascertained from wide-view TEM images, an example of which is shown in the inset of Fig. 1(e).

Finally, we used atomic force microscopy to quantify the roughness of Mn$_3$Sn. An example topographic map is shown in the inset of Fig. 1(d), which yields an average roughness of ≈ 0.5 nm over a 1 µm$^2$ region of a 50 nm Mn$_3$Sn (40$\bar{4}$3) film capped with 2 nm Ru.

Fig. 2(a) shows the magnetization (*M*) of a 60 nm Mn$_3$Sn (40$\bar{4}$3) film measured using superconducting quantum interference device vibrating sample magnetometry (SQUID-VSM, Quantum Design MPMS3), after subtracting the background contribution from a MgO substrate / 5 nm Ru reference sample. When magnetic field ($\mu_0 H$) is applied OP, with a component along the [2$\bar{1}$10] easy axis, an opening of the loop in the region ± 1.5 T is attributed to the reversal of the small uncompensated moment expected in the basal plane of Mn$_3$Sn.

With magnetic field applied along one of two orthogonal IP directions, a similar hysteretic behavior is observed, but with a smaller magnitude and an isotropic response. This is because magnetization is averaged over many crystallites, which can be four-fold symmetrically oriented with either [01$\bar{1}$0] (in the plane where uncompensated moment freely rotates) or partially IP [0001] (hard axis) directions parallel to the magnetic field. The inset of Fig. 2(a) plots magnetization response to magnetic field at different temperatures ($T$). At 5 K, a significant enhancement of magnetization demonstrates the appearance of the glassy FM phase [29].

X-ray magnetic circular dichroism (XMCD) was measured at the BESSY synchrotron facility [40], using negative, $\sigma_-$, polarized x-rays in an OP magnetic field of $\mu_0 H_\pm = \pm 8$ T. Non-zero XMCD around the Mn $L_{2,3}$ edges at 100 K, displayed in Fig. 2(b), confirms the presence of a net Mn moment that is reversible by external magnetic field. Using XMCD sum rule analysis [41], an uncompensated magnetic moment of $m_{s+l} = 0.279$ μB/f.u. was calculated, comprising of spin, $m_s = 0.273$ μB/f.u., and orbital, $m_l = 0.006$ μB/f.u., moments respectively. Measured $m_l$ is of the same order of magnitude as that simulated [25], whilst $m_s$ is found to be substantially larger than the theoretically predicted value for zero-field weak magnetization.

We explain this by considering the inset of Fig. 2(b), showing the magnetic field dependence of XMCD, measured as in Ref. [42]. XMCD increases approximately linearly, because of Mn spins tilting out of the film plane in response to strong applied magnetic fields [42]. This therefore enhances $m_s$, whose magnitude is in agreement with that determined from SQUID-VSM, which indicates similar paramagnetic behavior after the closing of the hysteresis loop. In addition, such SQUID-VSM measurements reveal that the remnant uncompensated moment within the basal plane is enhanced compared with bulk crystals. This may be exacerbated by structural defects and chemical disorder acting to modify the balance of AF exchange, DMI interactions and magnetocrystalline anisotropy that governs the spontaneous canting of Mn spins [39].

The inset of Fig. 2(b) also presents XMCD measured in a -8 T magnetic field at different temperatures. XMCD is reduced at 300 K, whilst its enhancement at 5 K evidences an increased net moment after transition to the spin glass state.

We now report on magnetotransport measurements in Mn$_3$Sn thin films lithographically patterned into 75 × 25 μm$^2$ Hall bar devices. Fig. 3(a) shows the variation in longitudinal resistivity ($\rho_{xx}$) for a 70 nm thick Mn$_3$Sn (0001) film as a function of temperature, during either zero-field cooling (ZFC) or cooling in a 7 T OP magnetic field (FC), followed in both cases by zero-field warming (ZFW). A background contribution from the 5 nm Ru buffer layer has been subtracted (see Supplemental Fig. S2). On cooling, a deviation from metallic behavior below 100 K, down to a bump in resistivity close to 50 K, provides evidence for the transition to the glassy FM phase. No change is observed between ZFC and FC protocols, as expected with cooling field parallel to the [0001] hard axis.

During subsequent ZFW, a thermal hysteresis is seen, with resistivity dipping between 50 K and 100 K. The resulting drop in resistivity is maintained until the film is warmed above room temperature. This may indicate that the transition from the inverse triangular to glassy FM state during cooling is not fully reversible until close to the Néel temperature ($T_N$ = 420 K [4]), for example due to pinning of the magnetic structure at AF domain walls discussed below.

Fig. 3(b) plots the variation in transverse resistivity ($\rho_{xy}$) as a function of OP magnetic field for the same 70 nm $Mn_3Sn$ (0001) film. At room temperature, we observe a linear Hall effect. This is because external field is applied along the $c$-axis, which is a magnetic hard axis, and is thus unable to manipulate the weak magnetization and induce a dominant chiral domain state yielding a net contribution to Berry curvature induced AHE. Between 100 K and 50 K, this ordinary Hall effect changes to an AHE as $Mn_3Sn$ transitions into the glassy FM phase. Here, moments cant spontaneously along the [0001] axis [30], with the resulting FM-like component of magnetization producing a conventional AHE in response to an OP magnetic field.

Moving onto the $Mn_3Sn$ ($40\bar{4}3$) films, we sourced current ($I_c$) along the [$01\bar{1}0$] direction and measured a component of Hall resistivity partially orthogonal to the (0001) crystal planes. Magnetic field was applied parallel to the [$40\bar{4}3$] direction, thus with a significant component along the [$2\bar{1}\bar{1}0$] easy axis. A schematic of this measurement configuration is inset in Fig. 4(b). These crystalline directions are as defined in one of the four equivalent structural grains (see Supplemental Fig. 5).

Figs. 4(a) and (b) show isothermal measurements of Hall effect for a 60 nm $Mn_3Sn$ ($40\bar{4}3$) film, in two temperature regimes. At room temperature, we measure a large

AHE driven by momentum-space Berry curvature [8]. A tiny longitudinal magnetoresistance (< 0.02 % at 8 T) is simultaneously observed (see Supplemental Material), ruling out FM contributions to magnetotransport.

Consistent AHE is measured across a series of $Mn_3Sn$ ($40\bar{4}3$) samples of different thicknesses, as discussed in the Supplemental Material. We extract a remnant anomalous Hall conductivity, $\sigma_{xy}$ ($\mu_0 H$ = 0 T) = 21 $\Omega^{-1}$ $cm^{-1}$, comparable to previous reports [35].

Fig. 4(a) shows that as temperature is increased towards $T_N$, coercive field decreases because of a softening of the magnetic structure. However, as temperature is decreased below 100 K, the magnitude of the Berry curvature generated AHE drops concomitantly, as $Mn_3Sn$ begins to leave the inverse triangular AF phase.

Below 50 K, $Mn_3Sn$ transitions into the glassy FM state, where momentum-space Berry curvature driven and (in this measurement configuration) conventional AHE are no longer detected. Instead, as shown in Fig. 4(b), the hysteresis loop becomes prominently asymmetric; a signature of the THE, caused when electrons acquire a real-space Berry phase upon encountering a magnetic structure with nonzero scalar spin chirality [31].

The inset of Fig. 4(b) explores this THE in a 30 nm $Mn_3Sn$ ($40\bar{4}3$) film undergoing different field cooling protocols. After ZFC following saturation in a 9 T OP magnetic field, a downward bump arising close to the point of magnetization reversal (when external field is swept from positive to negative) indicates a negative THE. A bump with the same polarity is seen on the reverse sweep. This demonstrates the formation of a noncollinear, noncoplanar spin texture close to coercivity, where AF order is reversing through domain propagation. Therefore, we attribute the THE to the nucleation of chiral domain walls. The even sign of THE suggests that the sense of rotation of spins in the domain walls is opposite during forward and reverse magnetization processes, whilst the internal moment configuration within the domain walls remains the same. A domain wall with such structure in the basal plane will appear as a noncollinear, noncoplanar spin texture possessing scalar spin chirality to electrons propagating within the plane of these Mn$_3$Sn (40$\bar{4}$3) films.

These results are in agreement with Li *et al* [34], who measure a symmetric planar Hall effect in Mn$_3$Sn single crystals, attributed to domain walls whose internal moment configuration is maintained during forward and reverse field sweeps (but whose sense of rotation is inverted). They also observe a dependence of the planar Hall effect polarity on applied field history.

We measure a similar magnetic state history in Mn$_3$Sn (40$\bar{4}$3) films. The inset of Fig. 4(b) shows that, when the thin film is ZFC after saturation in a -9 T OP magnetic field, the polarities of both symmetric THE peaks invert. This represents a sign change of real-space Berry phase, corresponding to an opposite sign of the finite scalar spin chirality generating it. We propose, therefore, that the handedness of the inverse triangular spin texture set at room temperature favors a certain chirality of domain wall formed during low temperature magnetization reversal.

Furthermore, the THE bumps with either negative or positive polarity are enhanced after cooling in a +9 T or -9 T OP magnetic field respectively. This may be explained by the external field further stabilizing the preferred domain wall moment configuration during cooling. We thus demonstrate a memory effect in noncollinear AF thin films, that can be controlled by setting the orientation of the inverse triangular spin texture from which the chirality of domain walls evolves, which may find applications in neuromorphic computing.

In conclusion, we have grown epitaxial thin films of Mn$_3$Sn with both (0001) *c*-axis orientation and (40$\bar{4}$3) crystallographic structure. In the latter case, Berry curvature driven AHE is observed at room temperature. Upon cooling through the magnetic phase transition at 50 K, a peak in the Hall resistivity indicates the appearance of a THE. The sign of this THE signal, and hence the chirality of the noncoplanar spin texture generating it, can be manipulated through cooling field conditions, thus furthering the potential of Mn$_3$Sn in *chiralitronic* devices.

**Acknowledgments**

We acknowledge Dr Dominik Kriegner for preparation of the Mn$_3$Sn (40$\bar{4}$3) $\chi$-$\phi$ XRD pole figure maps. We thank Prof Christian Back for provision of synchrotron beamtime at Helmholtz-Zentrum Berlin on beamline PM2 VEKMAG (under proposal 191-07803), and acknowledge Wolfgang Simeth and Denis Mettus for assistance with measurements. Financial support for the VEKMAG project and for the PM2 VEKMAG beamline are provided by the German Federal Ministry for Education and Research (BMBF 05K10PC2, 05K10WR1, 05K10KE1) and by Helmholtz-Zentrum Berlin, with Steffen Rudorff thanked for technical support. This work was partially funded by ASPIN (EU H2020 FET Open Grant 766566).

**Figure Captions**

FIG. 1 – (a) 2$\theta$-$\theta$ x-ray diffraction scans measured for a 70 nm Mn$_3$Sn (0001) film grown on a SrTiO$_3$ (111) substrate, and for a 60 nm Mn$_3$Sn (40$\bar{4}$3) film grown on a MgO (001) substrate. Inset shows azimuthal $\phi$ scans of the SrTiO$_3$ {202}, Ru {10$\bar{1}$1} and Mn$_3$Sn {20$\bar{2}$1} reflections for a 70 nm Mn$_3$Sn (0001) sample. The corresponding positions of in-plane crystallographic directions are indicated. (b) Cross-section scanning-TEM micrograph of a 70 nm Mn$_3$Sn (0001) film, viewed along the [$\bar{1}$10] zone axis of the SrTiO$_3$ (111) substrate. Inset shows a wide-view image over an extended region of the same lamella. (c) Representative crystal structure of Mn$_3$Sn (0001) films grown on SrTiO$_3$ (111) substrates using a Ru buffer layer. (d) $\chi$-$\phi$ x-ray pole figures measured for a 60 nm Mn$_3$Sn (40$\bar{4}$3) sample. The MgO [100] crystalline axis was aligned with $\phi$ = 0°. Inset shows an atomic force microscopy topographic map for a similar 50 nm Mn$_3$Sn (40$\bar{4}$3) film, capped with 2 nm Ru. (e) Cross-section high-resolution TEM micrograph of a 60 nm Mn$_3$Sn (40$\bar{4}$3) film, viewed along the [010] zone axis of the MgO (001) substrate. Inset shows an overview TEM image measured for the same sample. (f) Representative crystal structure of Mn$_3$Sn (40$\bar{4}$3) films grown on MgO (001) substrates using a Ru buffer layer.

FIG. 2 – (a) Magnetization measured as a function of magnetic field for a 60 nm $Mn_3Sn$ ($40\bar{4}3$) film at 300 K. Magnetic field was applied either out-of-plane or along one of two orthogonal in-plane directions (closed and open symbols represent down and up field sweeps respectively). Inset shows magnetization as a function of out-of-plane magnetic field, measured at different temperatures. (b) XAS and XMCD spectra for a 70 nm $Mn_3Sn$ ($40\bar{4}3$) film, recorded using $\sigma_-$ polarized x-rays in $\mu_0H_\pm = \pm 8$ T magnetic fields applied out-of-plane, at 100 K. Inset shows XMCD, calculated as the difference between spectra recorded with $\sigma_\pm$ polarized x-rays, measured in different out-of-plane magnetic fields at 100 K, as well as that measured in -8 T at 300 K and 5 K.

FIG. 3 – Magnetotransport for a 70 nm $Mn_3Sn$ (0001) film patterned into 75 × 25 µm² Hall bars, with 500 µA current parallel to the [$01\bar{1}0$] in-plane crystallographic direction. (a) Longitudinal resistivity measured as a function of temperature during either zero field cooling, or cooling in a 7 T magnetic field applied out-of-plane (closed symbols), and subsequent zero field warming (open symbols). Diagram illustrates measurement geometry in relation to different crystallographic directions in the $Mn_3Sn$ (0001) film. (b) Hall resistivity measured as a function of magnetic field applied out-of-plane at different temperatures. Transverse resistivity was measured along the [$\bar{2}110$] crystalline axis (closed and open symbols represent down and up field sweeps respectively).

FIG. 4 – Magnetotransport for a 60 nm $Mn_3Sn$ ($40\bar{4}3$) film patterned into 75 × 25 µm² Hall bars. Transverse resistivity was measured normal to the (0001) crystal plane, with 500 µA current parallel to the [$01\bar{1}0$] in-plane crystallographic direction (closed and open symbols represent down and up field sweeps respectively). Isothermal Hall resistivity measured as a function of magnetic field applied out-of-plane in the (a) higher temperature regime and (b) lower temperature regime. Diagram illustrates measurement geometry in relation to different crystallographic directions in the $Mn_3Sn$ ($40\bar{4}3$) film. Inset shows the Hall effect measured for a 30 nm $Mn_3Sn$ ($40\bar{4}3$) film at 5 K, after cooling under different field conditions.

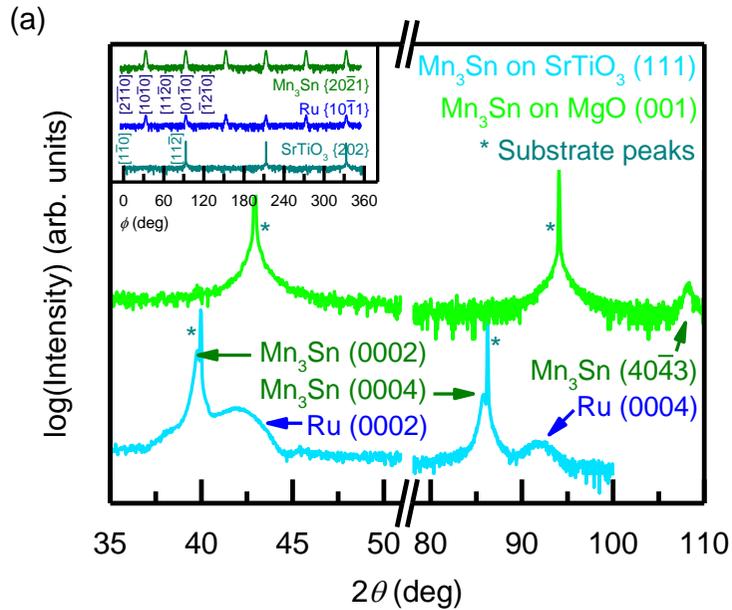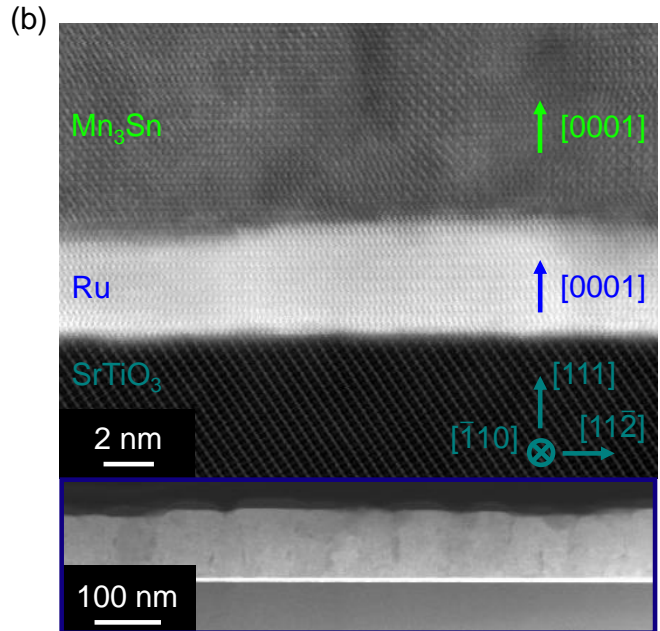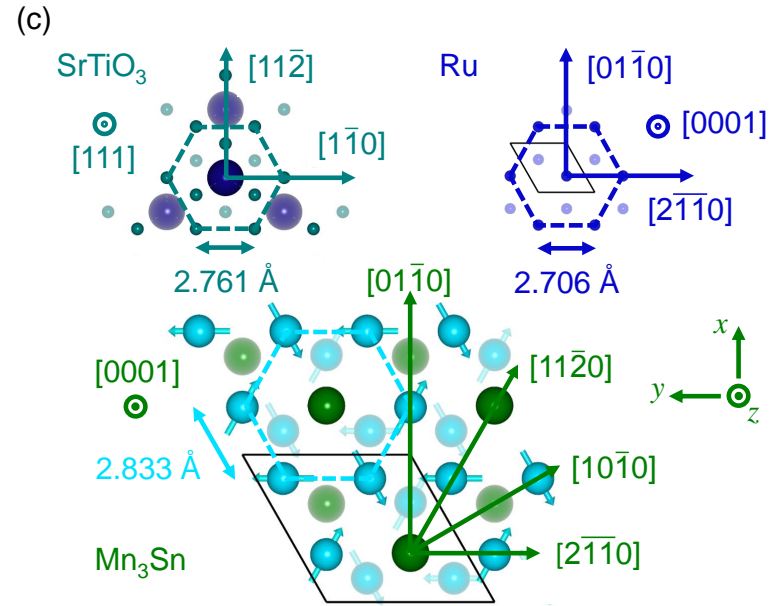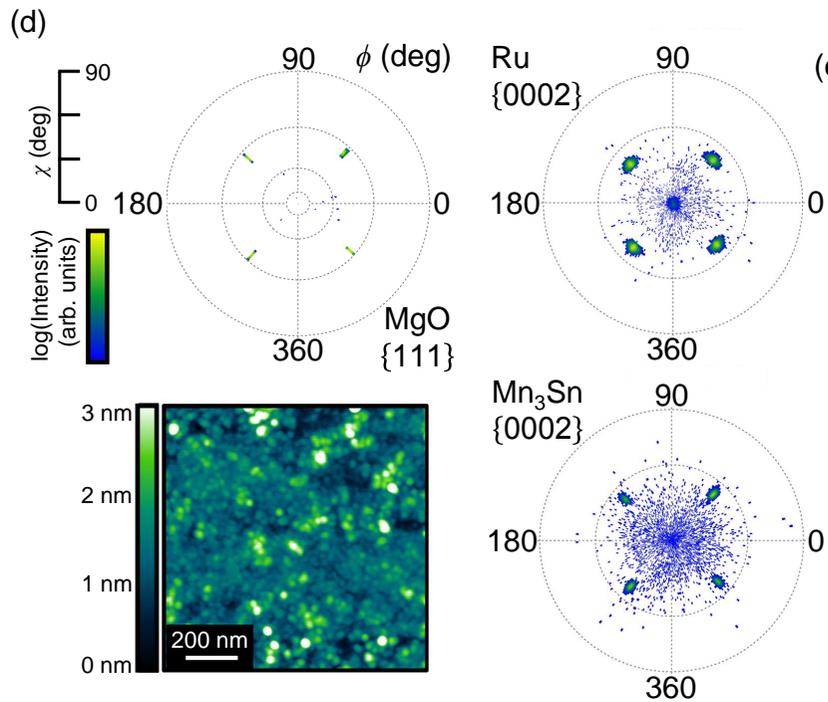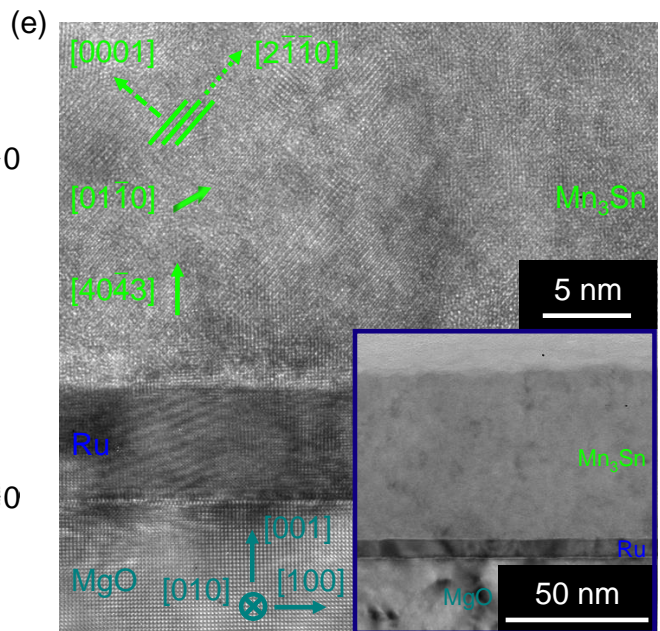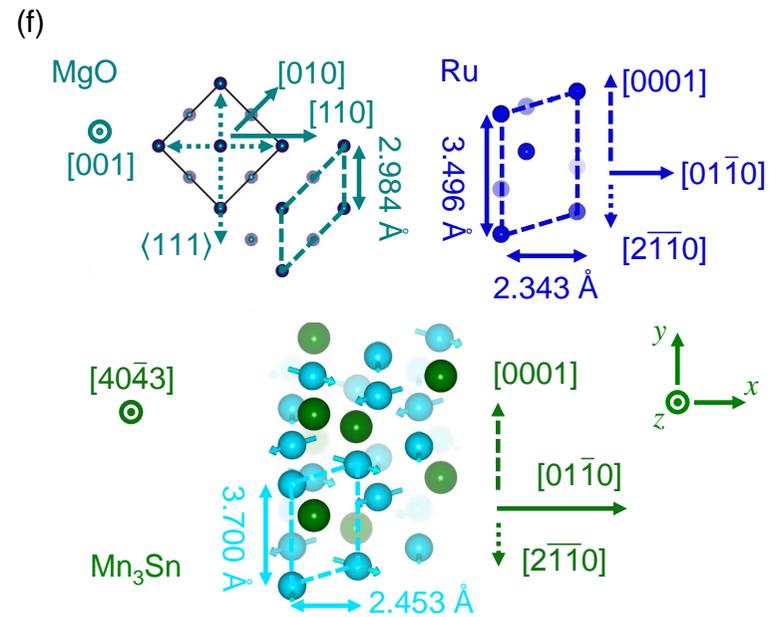

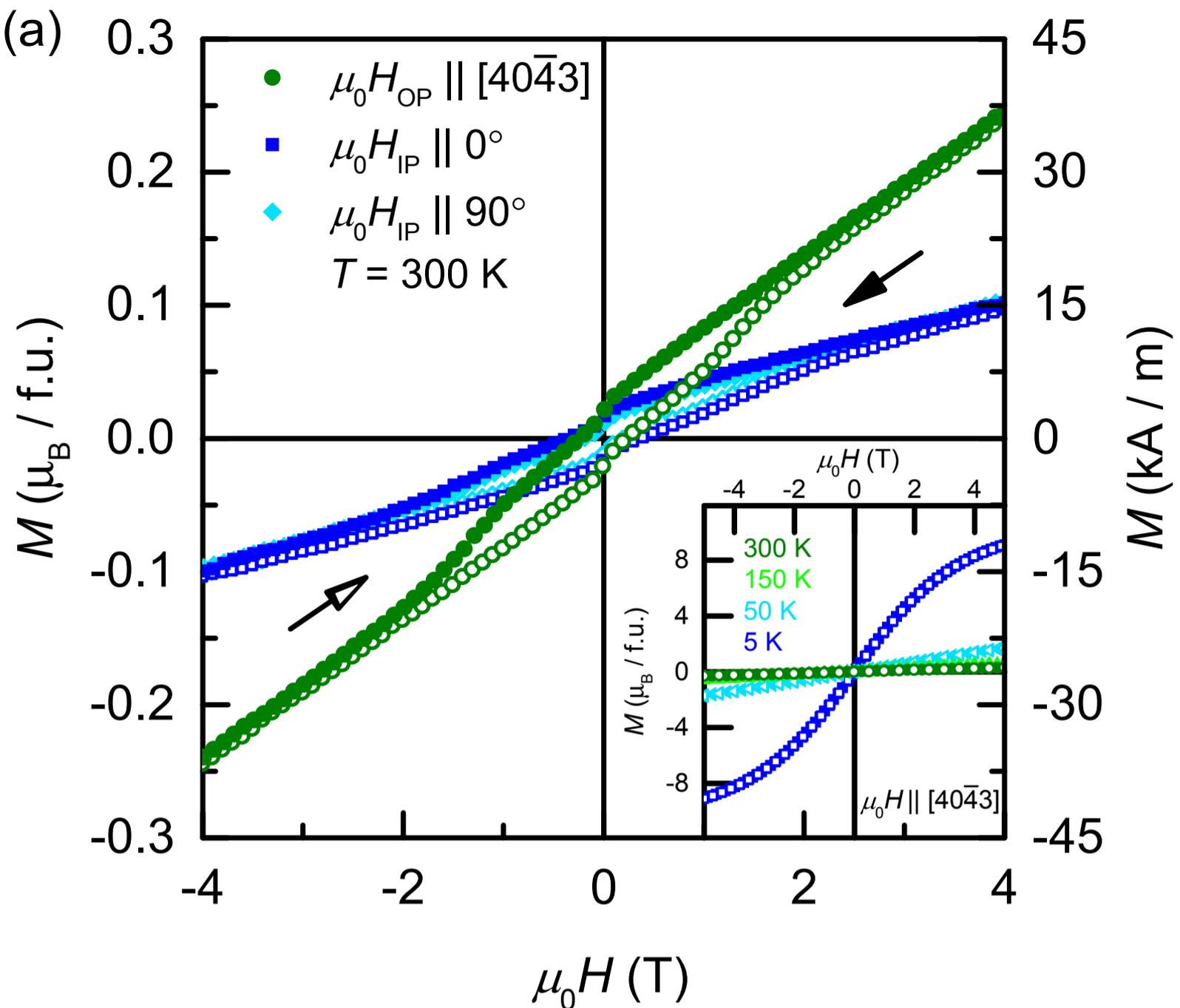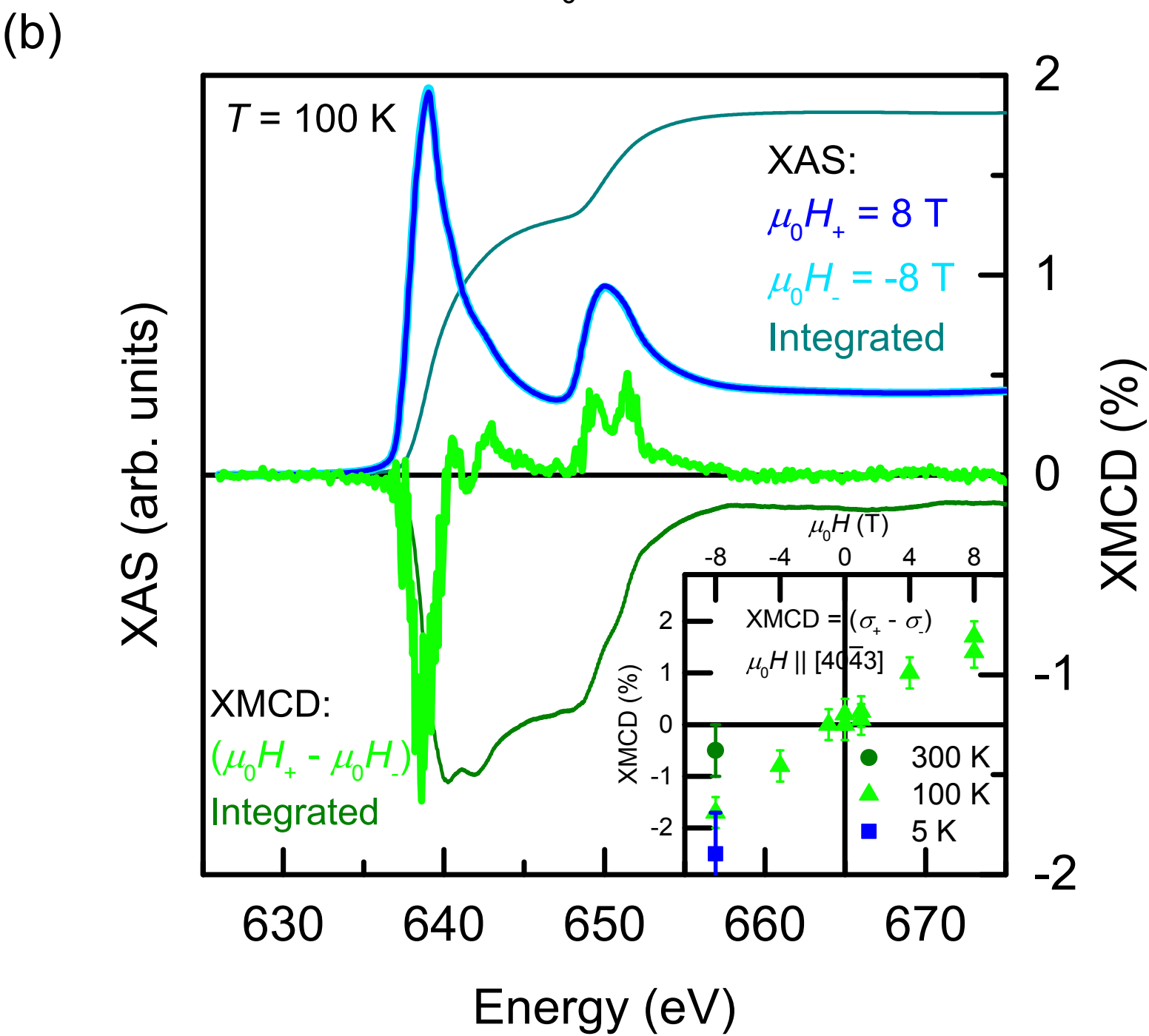

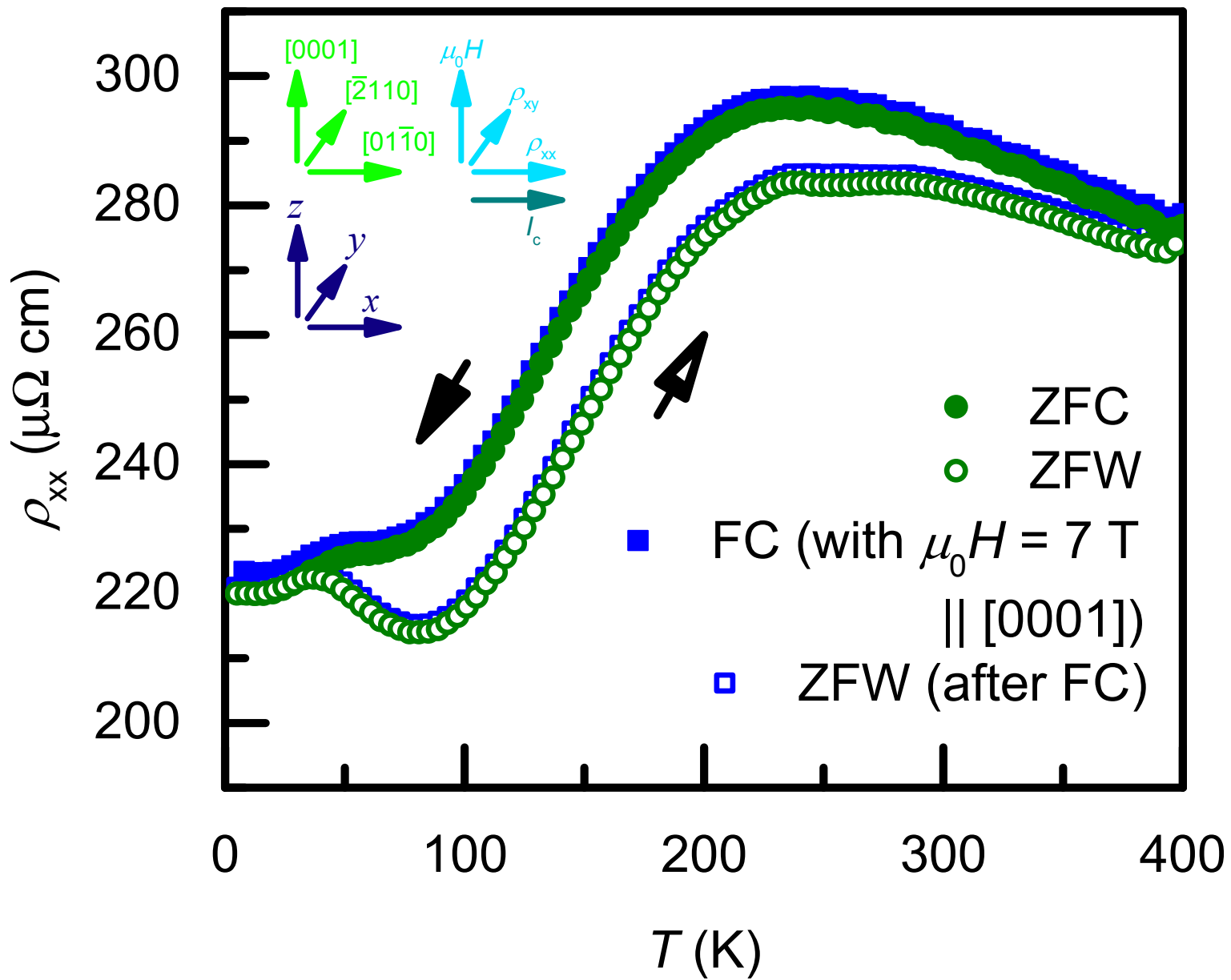

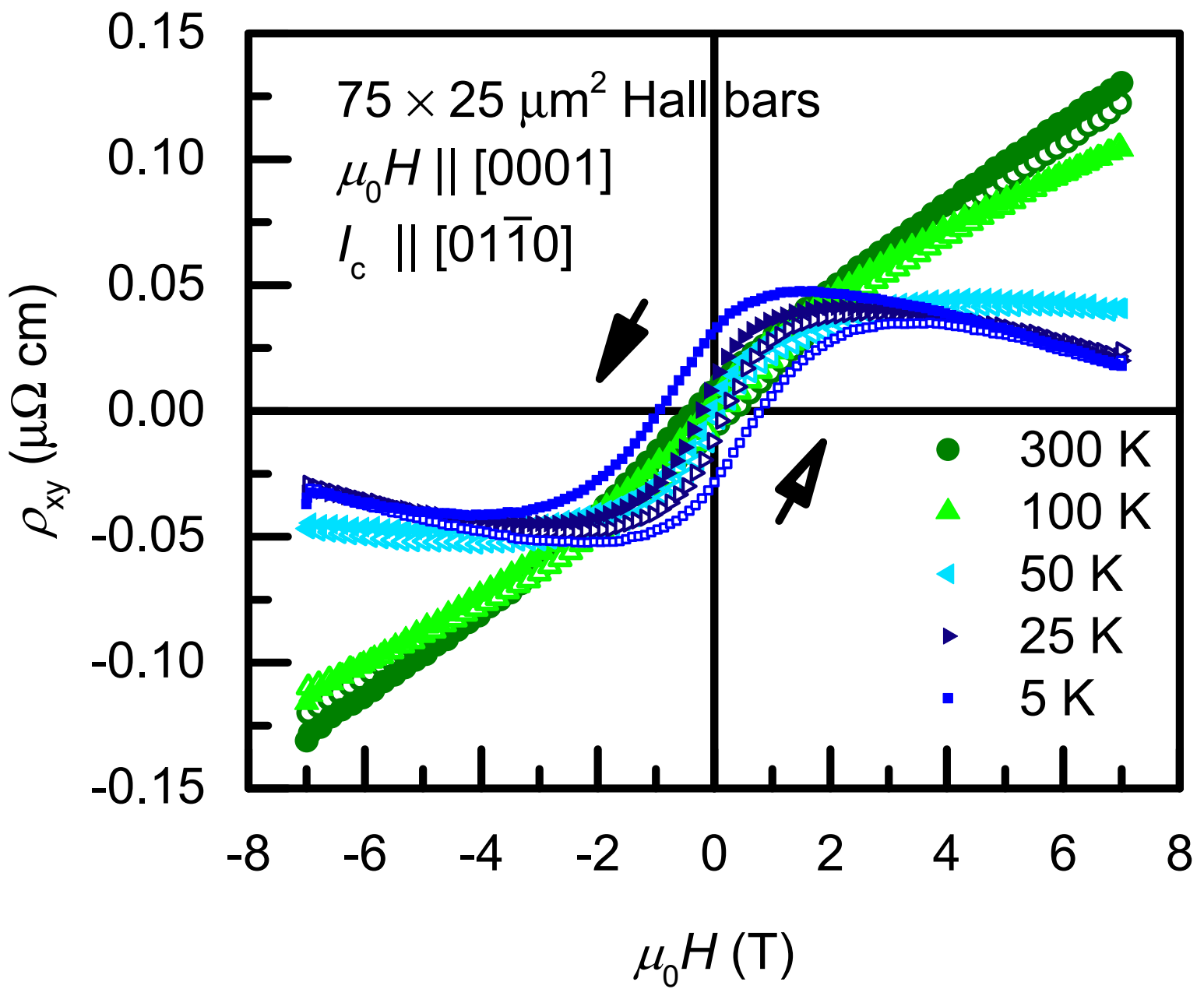

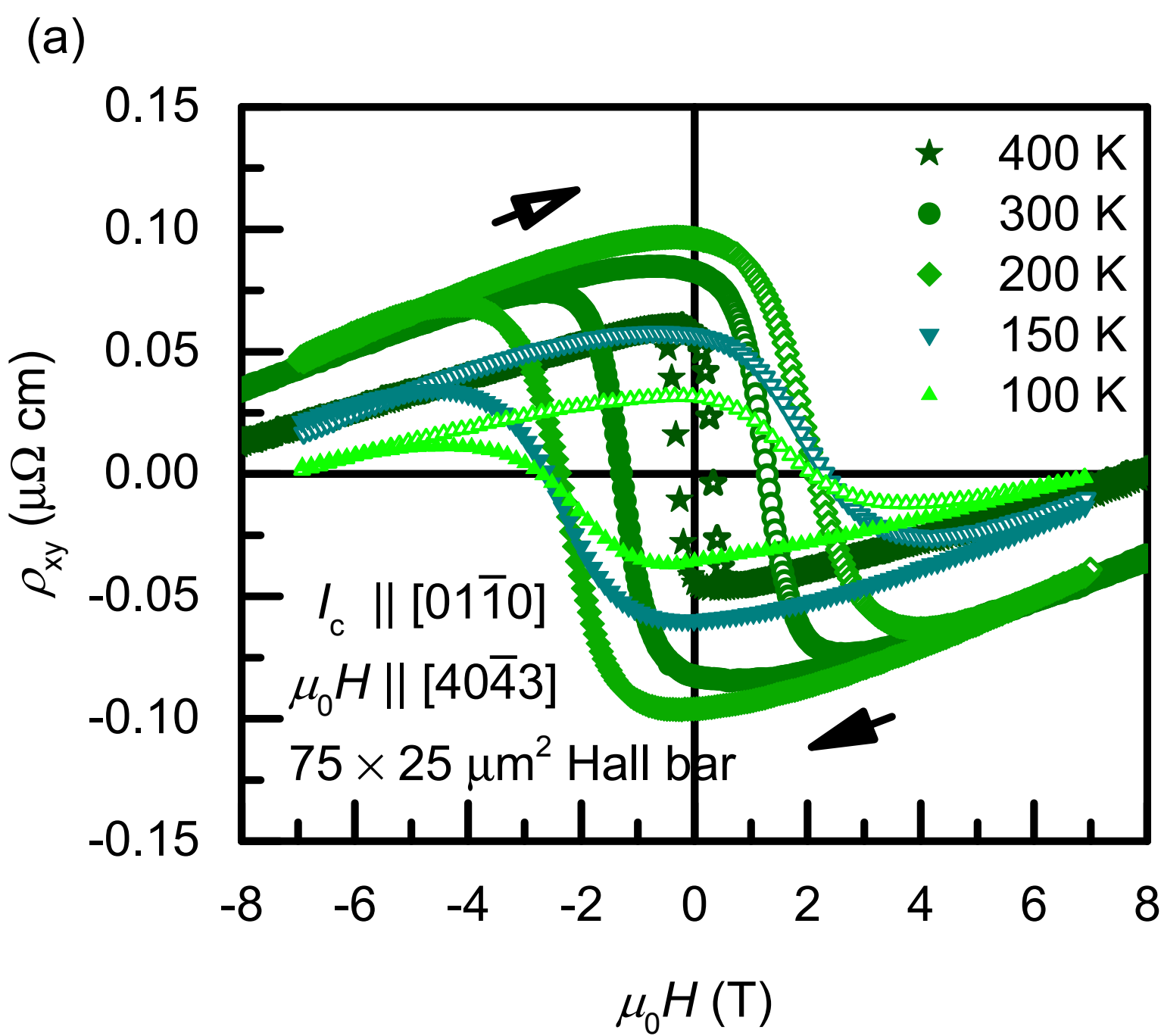
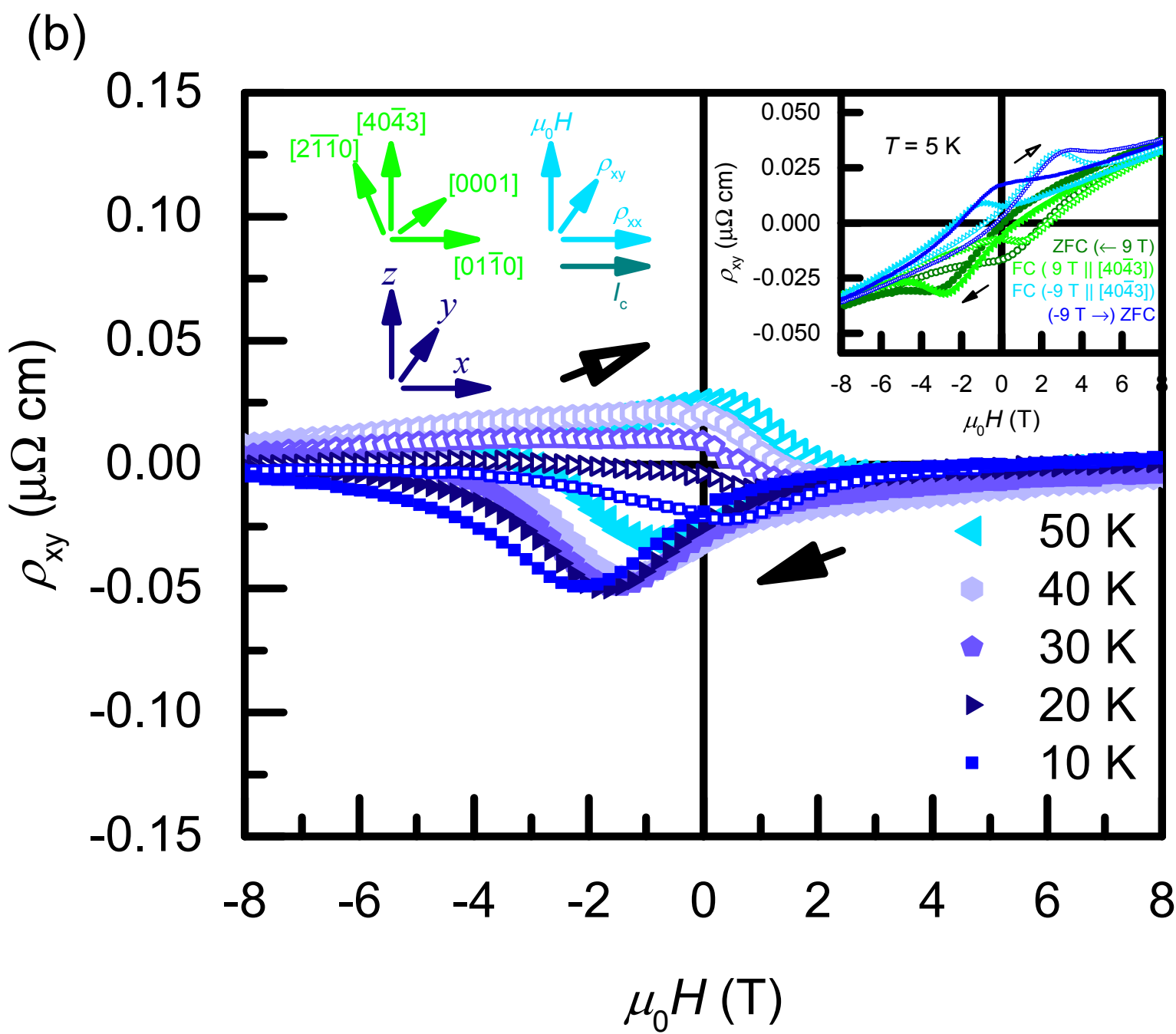

**Anomalous and topological Hall effects in epitaxial thin films of the noncollinear antiferromagnet Mn₃Sn**

James M. Taylor [*] [1], Anastasios Markou [2], Edouard Lesne [1], Pranava Keerthi Sivakumar [1], Chen Luo [3], Florin Radu [3], Peter Werner [1], Claudia Felser [2], Stuart S. P. Parkin [†] [1]

[1] Max Planck Institute of Microstructure Physics, Weinberg 2, 06120 Halle (Saale), Germany
[2] Max Planck Institute for Chemical Physics of Solids, Nöthnitzer Str. 40, 01187 Dresden, Germany
[3] Helmholtz-Zentrum Berlin for Materials and Energy, Albert Einstein Str. 15, 12489 Berlin, Germany

**Supplemental Material**

**I – Thin film growth and structural characterization**

The epitaxial thin films samples of Mn₃Sn utilized in this study were deposited using magnetron sputtering following the procedure described in Ref. [39] of the Main Text. However, in this case single crystal substrates of either (111) cut SrTiO₃ or (001) cut MgO were used to seed either (0001) *c*-axis texture or ($40\bar{4}3$) orientation respectively. In order to achieve continuous films, we made modifications to the temperatures during growth and post-annealing of the different layers. The Ru buffer layer was grown at 400°C, and allowed to cool to room temperature. Mn and Sn were then co-sputtered at room temperature, and the films post-annealed by heating to 300°C at a rate of 10°C / minute, holding for 10 minutes, then allowing to cool back to room temperature over a period of approximately 60 minutes.

Stacks were subsequently capped with 2.5 nm Al, which partially oxidizes, thus protecting the Mn₃Sn without shorting too much current during transport measurements. Since the surface of the resulting Al(Ox) is rough, samples used for AFM measurements were capped with 2 nm Ru, which follows the topography of the underlying Mn₃Sn closely, thus allowing an accurate quantification of its roughness.

---

[*] james.taylor@mpi-halle.mpg.de
[†] stuart.parkin@mpi-halle.mpg.de

Additional reference samples, consisting of either 30 nm or 5 nm Ru buffer layers alone, grown on MgO (001) and MgO (111) substrates respectively, were also prepared. Due to the proximity of the sets of peaks in 2$\theta$-$\theta$ position, the pole figure scans of the Ru ⟨0002⟩ reflections shown in Fig. 1(d) of the Main Text were measured from the 30 nm Ru buffer reference film. Meanwhile, the 5 nm reference sample was used to measure the contribution of the Ru to electrical transport, discussed below, and to record the background magnetization contribution from the substrate and buffer layer (subtracted from the total signal measured by SQUID-VSM, as explained in the discussion around Fig. 2(a) of the Main Text).

The composition of the thin film stacks was quantified, by energy dispersive x-ray spectroscopy measured in a scanning electron microscope, as $Mn_{0.76}Sn_{0.24}$. We denote this as $Mn_3Sn$, although we point out that the films grow with a slightly Mn-rich composition (and are thus not expected to show a first-order phase transition to a helical magnetic state below 275 K).

|  | Y:ZrO$_2$ (111) | SrTiO$_3$ (111) | Ru (0001) | Mn$_3$Sn (0001) | MgO (001) | Ru (40$\bar{4}$3) | Mn$_3$Sn (40$\bar{4}$3) |
|---|---|---|---|---|---|---|---|
| Y:ZrO$_2$ (111) | - |  | 5.7% | 10.7% |  |  |  |
| SrTiO$_3$ (111) |  | - | -2.0% | 2.6% |  |  |  |
| Ru (0001) | 5.7% | -2.0% | - | 4.7% |  |  |  |
| Mn$_3$Sn (0001) | 10.7% | 2.6% | 4.7% | - |  |  |  |
| MgO (001) |  |  |  |  | - | 17.2% | 24.0% |
| Ru (40$\bar{4}$3) |  |  |  |  | 17.2% | - | 5.8% |
| Mn$_3$Sn (40$\bar{4}$3) |  |  |  |  | 24.0% | 5.8% | - |

Mn$_3$Sn has a hexagonal crystal structure, with Mn moments arranged in a kagome lattice in the basal plane. Its bulk lattice parameters are: in-plane (IP), $a_{Mn3Sn}$ = 5.665 Å and out-of-plane (OP) $c_{Mn3Sn}$ = 4.531 Å. The bulk IP and OP lattice parameters of hexagonal Ru are $a_{Ru}$ = 2.706 Å and $c_{Ru}$ = 4.282 Å respectively. The lattice mismatches between the different orientations of the Mn$_3$Sn film and Ru buffer layer, and the various substrates (as well as the Y:ZrO$_2$ used in our earlier study [39]) are shown in Supplemental Table S1. The lattice mismatch between the SrTiO$_3$ (111) substrates and the Ru buffer layer (and, in turn, the subsequent c-axis oriented Mn$_3$Sn thin film) is less than that with the previously utilized Y:ZrO$_2$ substrates. This motivated our choice to switch to depositing on SrTiO$_3$ substrates. A smaller lattice mismatch likely results in a higher quality Ru buffer layer, which may subsequently contribute to the seeding of continuous Mn$_3$Sn films in the present case. The choice of cubic MgO (001) substrates, whilst poorly lattice matched with both Ru and Mn$_3$Sn, nevertheless succeeded in seeding the growth of an Ru buffer layer with sharp crystal structure, which is, in turn, relatively well lattice matched to the (40$\bar{4}$3) textured Mn$_3$Sn.

In both cases, x-ray diffraction (XRD) measurements were performed using a PANalytical X'Pert$^3$ diffractometer with Cu K$_{\alpha 1}$ radiation ($\lambda$ = 1.5406 Å). For the *c*-axis oriented films, by fitting the positions of the Mn$_3$Sn and Ru (0002) and (0004) peaks from the 2$\theta$-$\theta$ scan with a Gaussian profile, we can extract values for their OP hexagonal lattice parameters, *c*. These are found to be $c_{Mn_3Sn}$ = 4.524 Å and $c_{Ru}$ = 4.312 Å respectively. Separate measurements of the Mn$_3$Sn (20$\bar{2}$1) and Ru (10$\bar{1}$1) partially IP diffraction peaks (not shown) allow us to calculate the IP hexagonal lattice parameters, *a*. These are found to be $a_{Mn_3Sn}$ = 5.684 Å and $a_{Ru}$ = 2.695 Å respectively. In all cases, the measured lattice parameters are close to the bulk values, with no systematic epitaxial strain observed.

By setting the diffractometer to the 2$\theta$-$\theta$ position of each of these partially IP peaks and scanning the rotational angle, $\phi$, the azimuthal scans in the inset of Fig. 1(a) of the Main Text were recorded. The substrate was aligned with its [1$\bar{1}$0] edge along $\phi$ = 0°. From these, we determined the epitaxial relationship of the SrTiO$_3$ (111) / Ru (0001) / Mn$_3$Sn (0001) multilayers. In a similar way, the pole figures shown in Fig. 1(d) of the Main Text were mapped by performing individual $\phi$ azimuthal scans at consecutive values of the OP to IP rotation angle $\chi$.

We confirm this interpretation of the epitaxial relationship for the (40$\bar{4}$3) textured films through plane-view transmission electron microscopy (TEM) measurements. Supplemental Fig. S1 displays such a plane-view TEM image taken for a 60 nm Mn$_3$Sn (40$\bar{4}$3) film and viewed along the [00$\bar{1}$] MgO zone axis. The sample was prepared for measurement by conventional backside thinning and subsequent ion beam milling. The plane-view micrograph confirms the continuity of the films, whilst simultaneously demonstrating their granular structure. We observe large crystallites of around 200 nm in size, with grain boundary defects between then (which may contribute to enhancing the coercive field of these thin films through domain wall pinning, see below). TEM contrast is produced between neighboring crystallites exhibiting one of four possible orientations, with their c-axis aligned along one of the symmetric ⟨111⟩ directions in the cubic substrate, as discussed in the Main Text.

The selected area electron diffraction (SAED) pattern, shown in the right hand panel of Supplemental Fig. S1, supports this interpretation of the epitaxial growth mode. The ⟨01$\bar{1}$0⟩ reflections of the $Mn_3Sn$ film are coordinated with the cubic diagonal directions of the MgO, showing that the principle IP crystalline axis of $Mn_3Sn$ aligns with one of the four ⟨110⟩ directions in the substrate. Nevertheless, this SAED pattern confirms that individual grains possess a single-crystalline structure. The discrete, sharp diffraction spots demonstrate that the $Mn_3Sn$ film exhibits well defined IP crystallographic axes that are coherently oriented across multiple domains, with small relative mosaic spread between successive crystallites.

## II – Contribution of Ru buffer layer to electrical transport properties

The metallic Ru buffer layer will act to short some electric current during transport measurements. To quantify this contribution, the magnetotransport properties of a 5 nm Ru (0001) reference sample grown on MgO (111) were measured in the Van der Pauw geometry current applied parallel to the [2$\bar{1}\bar{1}$0] crystal direction. Electrical transport was measured in a Quantum Design Physical Property Measurement System (Dynacool PPMS), using a Keithley 6221 current source and 2182A nanovoltmeter. Supplemental Fig. S2 shows the variation in the longitudinal resistivity ($\rho_{xx}$) of the Ru layer as a function of temperature (*T*) during zero field cooling (ZFC) and zero field warming (ZFW). The resistivity decreases metallically, before saturating at low temperatures. This is because of defects and impurities modifying electrical transport in the low temperature regime.

Both longitudinal magnetoresistance (MR) and transverse resistivity ($\rho_{xy}$) were then measured as a function of magnetic field ($\mu_0H$) applied along the *c*-axis, shown in the upper and lower panels of the inset to Supplemental Fig. S2 respectively. We observe a Lorentz-type positive MR (driven by the cyclical motion of electrons in strong magnetic fields) and an ordinary Hall effect (with positive and negative gradients and high- and low-temperatures, respectively, showing a transition from hole- to electron-like carriers with decreasing temperature). Both these phenomena are as expected for a normal metal such as Ru, meaning that we can attribute any subsequently measured anomalous transport behavior to the $Mn_3Sn$ layer alone. In addition, the effects are small, particularly at room temperature, where $\rho_{xy}$ ($\mu_0 H$ = 8 T) = 0.03 $\mu\Omega$ cm. This, combined with the low thickness of the Ru (as compared with the $Mn_3Sn$ films), means that the buffer layer will make a minimum contribution to the overall magnetotransport behavior of the full stack.

### III – Thickness dependence of AHE in $Mn_3Sn$ ($40\bar{4}3$) thin films

To demonstrate the negligible magnetotransport contribution of the Ru layer explicitly, in Supplemental Fig. S3 we plot the Hall effect of MgO (001) substrate / 5 nm Ru buffer / $Mn_3Sn$ ($40\bar{4}3$) films after the subtraction of the ordinary Hall effect background measured from the Ru reference sample, using a parallel resistors model. The resulting Hall effect of the $Mn_3Sn$ layer alone does not differ qualitatively, beside the partial removal of a linear background, whilst the minor quantitative change reflects the small role of the Ru buffer in the magnetotransport behavior of the overall multilayer. Here the samples were measured in the Van der Pauw geometry, although the results are in very good agreement with those measured in patterned Hall bar devices (where the contribution from the buffer layer is not routinely subtracted), as reported in the Main Text.

Transverse resistivity was measured at 300 K as a function of magnetic field applied OP for $Mn_3Sn$ ($40\bar{4}3$) films of different thicknesses in the range 70 nm down to 30 nm. All films exhibit a large Berry curvature driven AHE, as discussed in the Main Text. Although the magnitude of Hall resistivity decreases as thickness decreases, we otherwise observe no qualitative change in the AHE generated by the topology of the noncollinear AF structure. In particular, the coercive field, which reflects the reversal of the handedness of the inverse triangular AF spin texture through a mechanism of chiral domain nucleation and propagation (and is enhanced by the pinning of AF domain walls at crystallite grain boundaries as discussed above), remains the same across the thickness series.

In order to compare this thickness dependence of AHE directly with previous studies, we calculate Hall conductivity ($\sigma_{xy} = \rho_{xy} / \rho_{xx}^2$) for the Mn$_3$Sn (40$\bar{4}$3) films of different thickness, using their simultaneously measured longitudinal resistivity, as plotted in the right-hand panel of Supplemental Fig. S3. The Hall conductivity remains broadly the same across the thickness series, with a small decrease observed for the 30 nm film, which may reflect a reduction in crystal structure quality at such low thickness. We extract a typical remnant Hall conductivity of $\sigma_{xy}$ ($\mu_0 H$ = 0 T) = 21 $\Omega^{-1}$ cm$^{-1}$, which, whilst small than that measured for bulk single crystals [8], is larger than the planar Hall effect measured for epitaxial Mn$_3$Sn films [38], and is comparable to values measured for polycrystalline bulk [33] or thin film [35] samples.

Such large measured values of anomalous Hall conductivity, which persist even at zero magnetic field (thus reflecting the stability of the chiral domain structure even with external magnetic field removed), demonstrate the suitability of these noncollinear AF Mn$_3$Sn films for topological spintronic applications. Comparison of the transverse and longitudinal conductivities of our thin films confirms that they lie in the regime where intrinsic AHE dominates, comparable to other materials shown to demonstrate Berry curvature driven topological transport properties.

## IV – Further magnetotransport properties of Mn$_3$Sn (0001) thin films

Supplemental Fig. S4 shows further temperature dependent measurements of the Hall effect in Mn$_3$Sn (0001) films. In this case, we measure transverse resistivity || [$\bar{1}$100], as a function of magnetic field || [0001], for a 75 ×25 µm$^2$ Hall bar fabricated with its long axis parallel to the [11$\bar{2}$0] crystallographic direction. In systems with hexagonal structure, the [11$\bar{2}$0] crystalline axis is distinct from the [01$\bar{1}$0] direction (as probed in Fig. 3(b) of the Main Text). However, we see that an almost identical Hall effect is measured with current flowing along both crystal directions, with very similar magnitude and temperature dependence.

This can be explained by considering the azimuthal XRD ϕ scans presented in Fig. 1(a) of the Main Text. We observe a six-fold symmetry of the partially IP Mn$_3$Sn {20$\bar{2}$1} reflections (as opposed to the expected three-fold recurrence), suggesting the hexagonal crystal structure exhibits rotational twinning between basal planes. This rotational twinning of Mn$_3$Sn will cause magnetotransport to average over crystallographic directions at both 30° and 60° intervals, thus resulting in an identical Hall effect along both the ⟨01$\bar{1}$0⟩ and ⟨2$\bar{1}\bar{1}$0⟩ crystalline axes.

The inset of Supplemental Fig. S4 shows longitudinal MR, measured ∥ [01$\bar{1}$0] with external magnetic field applied along the *c*-axis. We observe an extremely small MR at 300 K, indicating an absence of net magnetization in these antiferromagnetic thin films. As temperature is decreased to 5 K, MR increases as Mn$_3$Sn transitions into the glassy FM state. This may be the result of increased scattering of charge carriers from the frustrated magnetic moments in the spin glass phase.

**V – Further magnetotransport properties of Mn$_3$Sn (40$\bar{4}$3) thin films**

Supplemental Fig. S5 shows further temperature dependent measurements of the Hall effect in 60 nm Mn$_3$Sn (40$\bar{4}$3) films. This time, we measure transverse resistivity ∥ [01$\bar{1}$0], with current applied in a direction partially perpendicular to consecutive (0001) crystal planes (in a geometry orthogonal to that probed in Fig. 4 of the Main Text).  However, we observe a very similar magnitude and temperature dependence of Hall effect in both configurations.

This is to be expected, when we consider the four-fold symmetry of crystal grains discussed in the Main Text. A given mesoscopic device will consist of multiple crystallites, each of which will have either a ⟨01$\bar{1}$0⟩ direction, or the *c*-axis, parallel to current flow, depending on which of the four possible cubic axes of the substrate that grain follows. Thus, for a mesoscopic device, it is not possible to distinguish between the IP [01$\bar{1}$0] and partially IP [0001] directions across multiple grains. Hall resistivity is averaged over all crystallites (although both measurement configurations should yield equally large Hall effect [7]), thus resulting in an isotropic Hall effect along both the ⟨01$\bar{1}$0⟩ and ⟨0001⟩ crystalline axes.

The coercive field is 1.3 T, close to the reversal field of the inverse triangular AF order, indicating the sign of the AHE depends on the chirality of the spin texture. This coercivity is comparable to that reported elsewhere in the literature for $Mn_3Sn$ thin films [35-37], but is higher than that measured for bulk single crystal samples. This can be explained by the structural defects inherent in such thin film samples pinning AF domain wall motion during the reorientation process of the inverse triangular spin texture, and thus indicates the key role played by chiral domains dynamics in determining the magnetotransport behavior of topological antiferromagnets.

We also note that, as temperature is decreased between 300 K and 50 K, the square shape of the AHE hysteresis loop remains unchanged. There is no evidence of a transition to a helical magnetic phase [28] in these $Mn_3Sn$ ($40\bar{4}3$) films, as expected for samples that are grown with a slight excess of Mn.

The inset of Supplemental Fig. S5 shows longitudinal MR || [$01\bar{1}0$]. As discussed in the main text, we observe an extremely small MR at 300 K, reflecting the tiny uncompensated magnetic moment and ruling out ferromagnetism as the origin of the large room temperature AHE observed in these thin films.

Very similar behavior of the Hall effect in $Mn_3Sn$ ($40\bar{4}3$), specifically a large AHE at room temperature and a transition to THE below 50 K, has been observed across a selection of films with comparable crystal quality but different thickness, and grown using an identical technique but in different 'batches'. This gives us confidence that the effects observed are general for this material, and are reproducible.

Temperature dependent magnetotransport properties were studied in such samples of different thicknesses after patterning into Hall bar devices, with the example of a 30 nm $Mn_3Sn$ ($40\bar{4}3$) film presented in Supplemental Fig. S6. Hall resistivity has a lower magnitude than in thicker films, as discussed above. In addition, Supplemental Fig. S6 shows a steeper linear background in the high magnetic field region. This is because of the lower thickness of the $Mn_3Sn$ film as compared with the Ru buffer layer, whose ordinary Hall contribution has not be subtracted in this case. Nevertheless, the temperature dependence of the Hall effect is in very good agreement with the other samples in the thickness series. This demonstrates that, down to 30 nm, there is little thickness dependence in the magnetotransport properties of $Mn_3Sn$.

The inset of Supplemental Fig. S6 shows longitudinal MR, measured ∥ [01$\bar{1}$0] with magnetic field applied OP. Again, we observe a tiny MR at 300 K, reflecting an absence of net magnetization in these $Mn_3Sn$ (40$\bar{4}$3) thin films and thus excluding ferromagnetism as the origin of the AHE observed at room temperature. As temperature is decreased to 5 K, MR increases as the material transition into the glassy FM phase.

A further example is presented in Supplemental Fig. S7. Here, we measure Hall effect at 300 K in a 50 nm $Mn_3Sn$ (40$\bar{4}$3) film, showing momentum-space Berry curvature driven AHE, and at 25 K in a 40 nm $Mn_3Sn$ (40$\bar{4}$3) film, showing a real-space Berry curvature driven THE. In the latter case, the sample has been ZFC from 400 K, after saturation in a 9 T external magnetic field.

The inset of Supplemental Fig. S7 shows an optical image of a typical patterned Hall bar, with the measurement geometry indicated. Devices were fabricated using a combination of electron beam lithography and Ar ion etching, with sizes ranging from 75 × 25 µm$^2$ down to 3 × 1 µm$^2$. Different sized Hall bars were, in turn, patterned with their long axis (direction of current flow) along either the [01$\bar{1}$0] crystallographic axis, or a direction parallel to the IP component of the [0001] *c*-axis, as well as at 30° intervals in between.

The left-hand panel of Supplemental Fig. S12 shows Hall effect measured at room temperature for a 30 nm $Mn_3Sn$ (40$\bar{4}$3) film, in 75 × 25 µm$^2$ Hall bars fabricated along the [01$\bar{1}$0] crystalline axis, as well as at 30° and at 60° to this direction. We observe a very similar Hall effect for current flow along all crystallographic directions. This can be explained by the fact that the weak magnetization in the basal plane of $Mn_3Sn$, caused by the canting of Mn moments towards the direction joining nearest non-magnetic neighbor atoms, is free to rotate in response to an external magnetic field [5]. Therefore, regardless of the measurement geometry, external magnetic field will corresponding orient the chirality of the inverse triangular spin texture, resulting in isotropic AHE with respect to the hexagonal crystal lattice.

Finally, the right-hand panel of Supplemental Fig. S12 shows Hall effect measured at room temperature for a 60 nm $Mn_3Sn$ ($40\bar{4}3$) film, in Hall bars of different dimensions fabricated along the $[01\bar{1}0]$ crystallographic direction. Again, an almost identical AHE, of equal magnitude and with the same coercive field, is observed in devices down to 3 × 1 µm$^2$ in size. Because we would expect a modification of Hall effect upon approach to measurement of a single domain state, this suggests that the length scale of chiral domains in $Mn_3Sn$ is less than 1 µm, which would indeed be the case if AF domain dimensions are correlated with the size of crystallite grains in the ($40\bar{4}3$) thin films.

**Figure Captions**

Supplemental Table S1 – Lattice mismatch calculated between different orientations of Mn$_3$Sn thin films and Ru buffer layers, and the various substrates utilized to achieve each epitaxial relationships. We calculate lattice mismatch from bulk lattice parameters, using the same layer stacking sequences as in our deposited multilayers.

Supplemental FIG. S1 – Plane-view TEM image of a 60 nm Mn$_3$Sn (40$\bar{4}$3) film, viewed along the [00$\bar{1}$] zone axis of the MgO (001) substrate. The primary [01$\bar{1}$0] crystallographic direction of the Mn$_3$Sn layer lies in-plane, with the [0001] and [2$\bar{1}\bar{1}$0] crystalline axes partially out-of-plane. The coordinate system defined for subsequent magnetotransport measurements is indicated. Right panel shows a selected-area electron diffraction pattern measured for the same planar sample, with reflections from the substrate and Mn$_3$Sn thin film indexed.

Supplemental FIG. S2 – Magnetotansport for a 5 nm Ru (0001) thin film in the van der Pauw geometry, with 500 µA current parallel to the in-plane [2$\bar{1}\bar{1}$0] crystallographic direction: Longitudinal resistivity measured as a function of temperature during zero field cooling (closed symbols) and subsequent zero field warming (open symbols). Longitudinal magnetoresistance (upper inset) and Hall resistivity (lower inset) measured at different temperatures as a function of external magnetic field applied out-of-plane. Transverse resistivity was measured along the [01$\bar{1}$0] crystalline axis (closed and open symbols represent down and up field sweeps respectively).

Supplemental FIG. S3 – Hall effect measured at 300 K as a function of external magnetic field applied out-of-plane for a different thicknesses of Mn$_3$Sn (40$\bar{4}$3) films in the van der Pauw geometry. A background contribution, as recorded from a 5 nm Ru buffer layer reference sample, has been subtracted. Left panel plots transverse resistivity, measured normal to the (0001) crystal plane, with 500 µA current parallel to the [01$\bar{1}$0] in-plane crystallographic direction (closed and open symbols represent down and up field sweeps respectively). Right panel plots the same measurements converted to Hall conductivity, as calculated using the longitudinal resistivity of each film.

Supplemental FIG. S4 – Hall resistivity measured at different temperatures as a function of external magnetic field applied out-of-plane for a 70 nm Mn$_3$Sn (0001) film patterned into a 75 × 25 µm$^2$ Hall bar. Transverse resistivity was measured along the [$\bar{1}$100] crystalline axis, with 500 µA current parallel to the [11$\bar{2}$0] in-plane crystallographic direction (closed and open symbols represent down and up field sweeps respectively). Inset shows longitudinal

magnetoresistance, parallel to the [01$\bar{1}$0] crystal direction, measured over the same out-of-plane magnetic field range at different temperatures.

Supplemental FIG. S5 – Hall resistivity measured at different temperatures as a function of external magnetic field applied out-of-plane, for a 60 nm $Mn_3Sn$ (40$\bar{4}$3) film patterned into a 75 × 25 µm$^2$ Hall bar. Transverse resistivity was measured along the [0$\bar{1}$10] in-plane crystallographic direction, with 500 µA current normal to the (0001) crystal plane (closed and open symbols represent down and up field sweeps respectively). Inset shows longitudinal magnetoresistance, parallel to the [01$\bar{1}$0] crystal direction, measured over the same out-of-plane magnetic field range at 300 K.

Supplemental FIG. S6 – Hall resistivity measured at different temperatures as a function of external magnetic field applied out-of-plane, for a 30 nm $Mn_3Sn$ (40$\bar{4}$3) film patterned into a 75 × 25 µm$^2$ Hall bar. Transverse resistivity was measured normal to the (0001) crystal plane, with 500 µA current parallel to the [01$\bar{1}$0] in-plane crystallographic direction (closed and open symbols represent down and up field sweeps respectively). Inset shows longitudinal magnetoresistance, parallel to the [01$\bar{1}$0] crystal direction, measured over the same out-of-plane magnetic field range at different temperatures.

Supplemental FIG. S7 – Hall resistivity measured as a function of external magnetic field applied out-of-plane, at 300 K for a 50 nm $Mn_3Sn$ (40$\bar{4}$3) film and at 25 K for a 40 nm $Mn_3Sn$ (40$\bar{4}$3) film, patterned into 75 × 25 µm$^2$ Hall bars. Transverse resistivity was measured normal to the (0001) crystal plane, with 500 µA current parallel to the [01$\bar{1}$0] in-plane crystallographic direction (closed and open symbols represent down and up field sweeps respectively). Inset shows optical image of a typical device, with 75 × 25 µm$^2$ Hall bars fabricated along different crystalline axes and measurement geometry indicated.

Supplemental FIG. S8 – Hall resistivity measured at 300 K as a function of external magnetic field applied out-of-plane, with 500 µA current along the Hall bar long axis (closed and open symbols represent down and up field sweeps respectively): Left frame shows Hall effect for a 30 nm $Mn_3Sn$ (40$\bar{4}$3) film patterned into 75 × 25 µm$^2$ Hall bars directed along 30° intervals to the [01$\bar{1}$0] crystalline axis. Right frame shows Hall effect for a 60 nm $Mn_3Sn$ (40$\bar{4}$3) film patterned into Hall bars of different sizes parallel to the [01$\bar{1}$0] in-plane crystallographic direction.

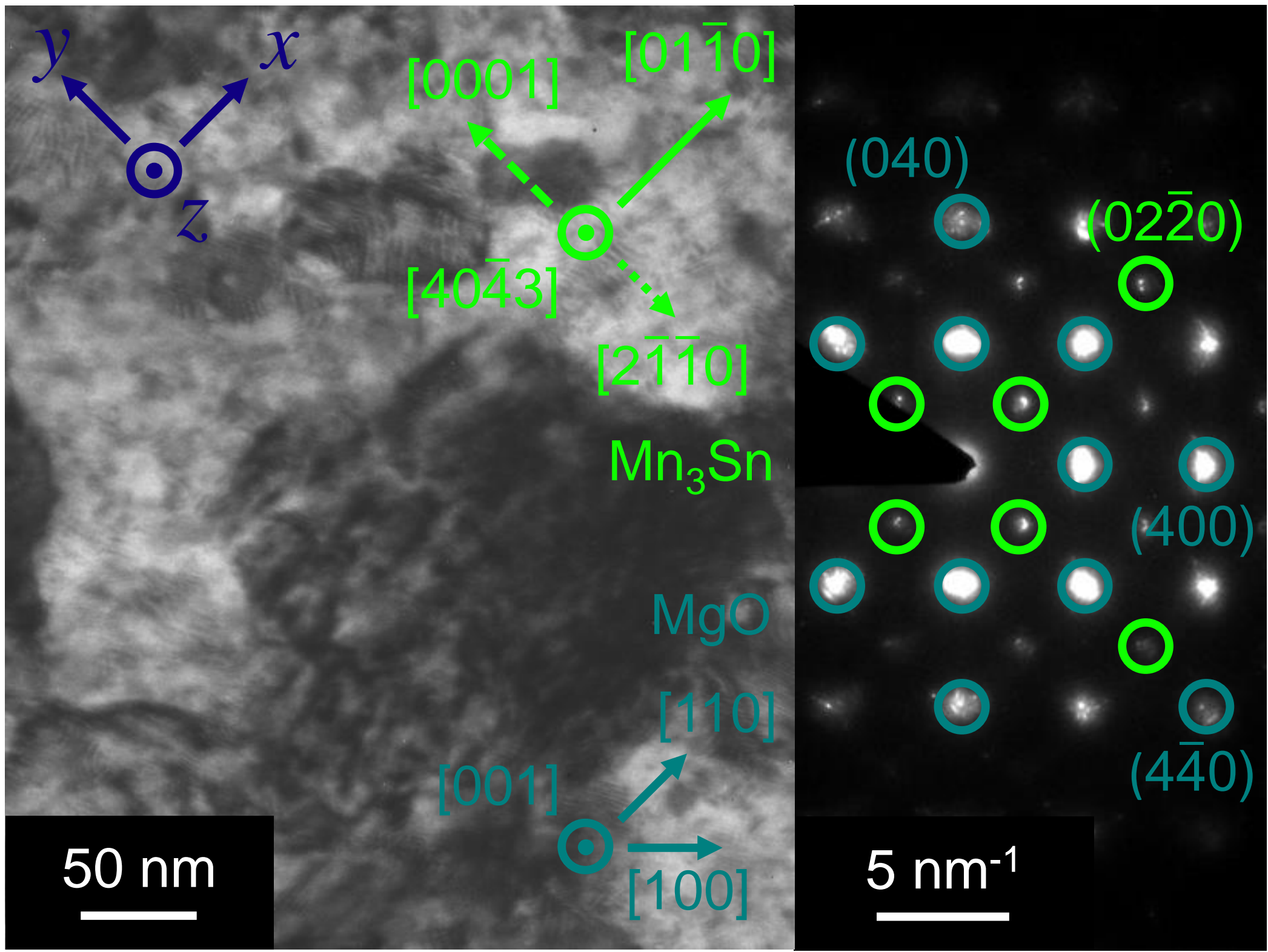

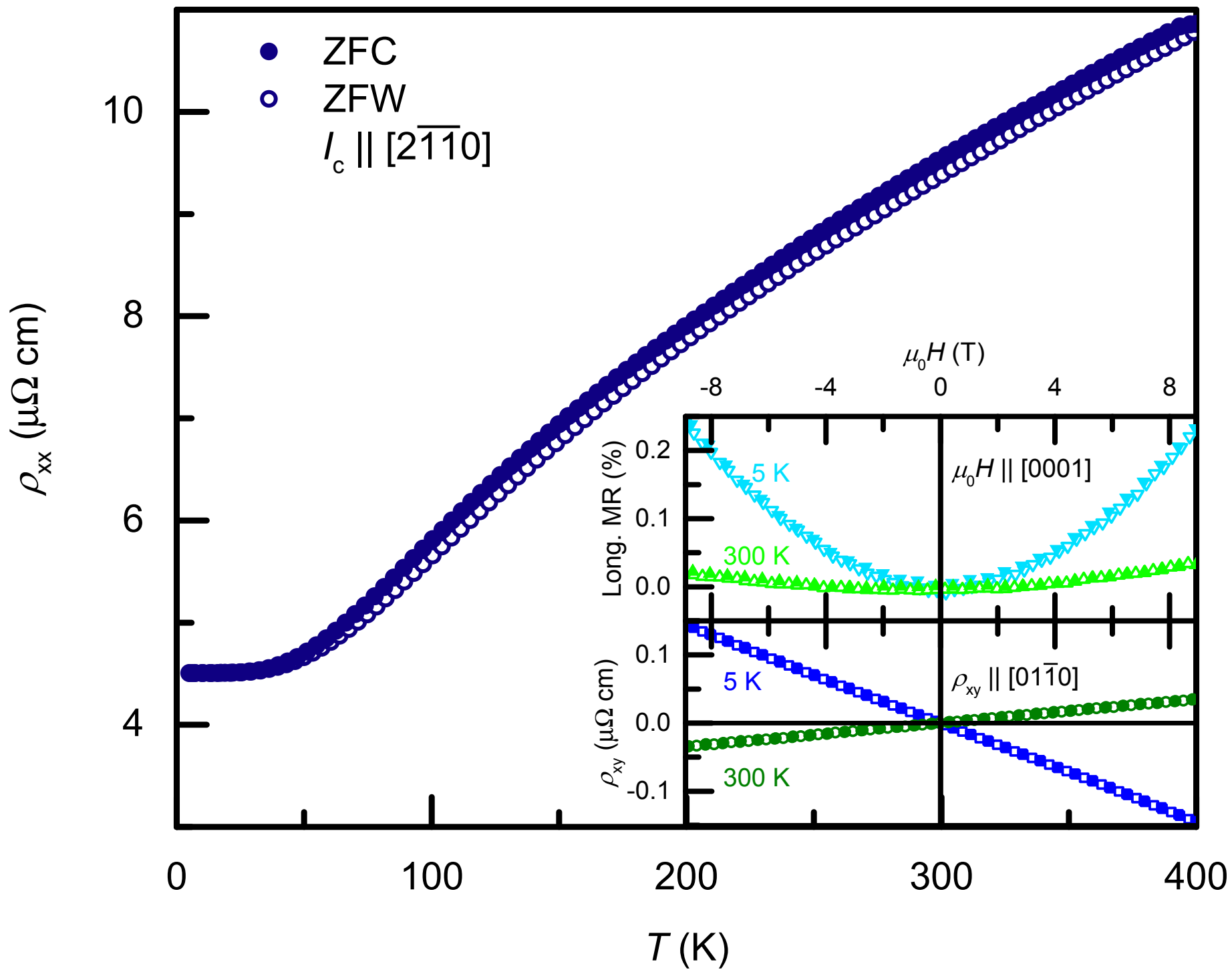

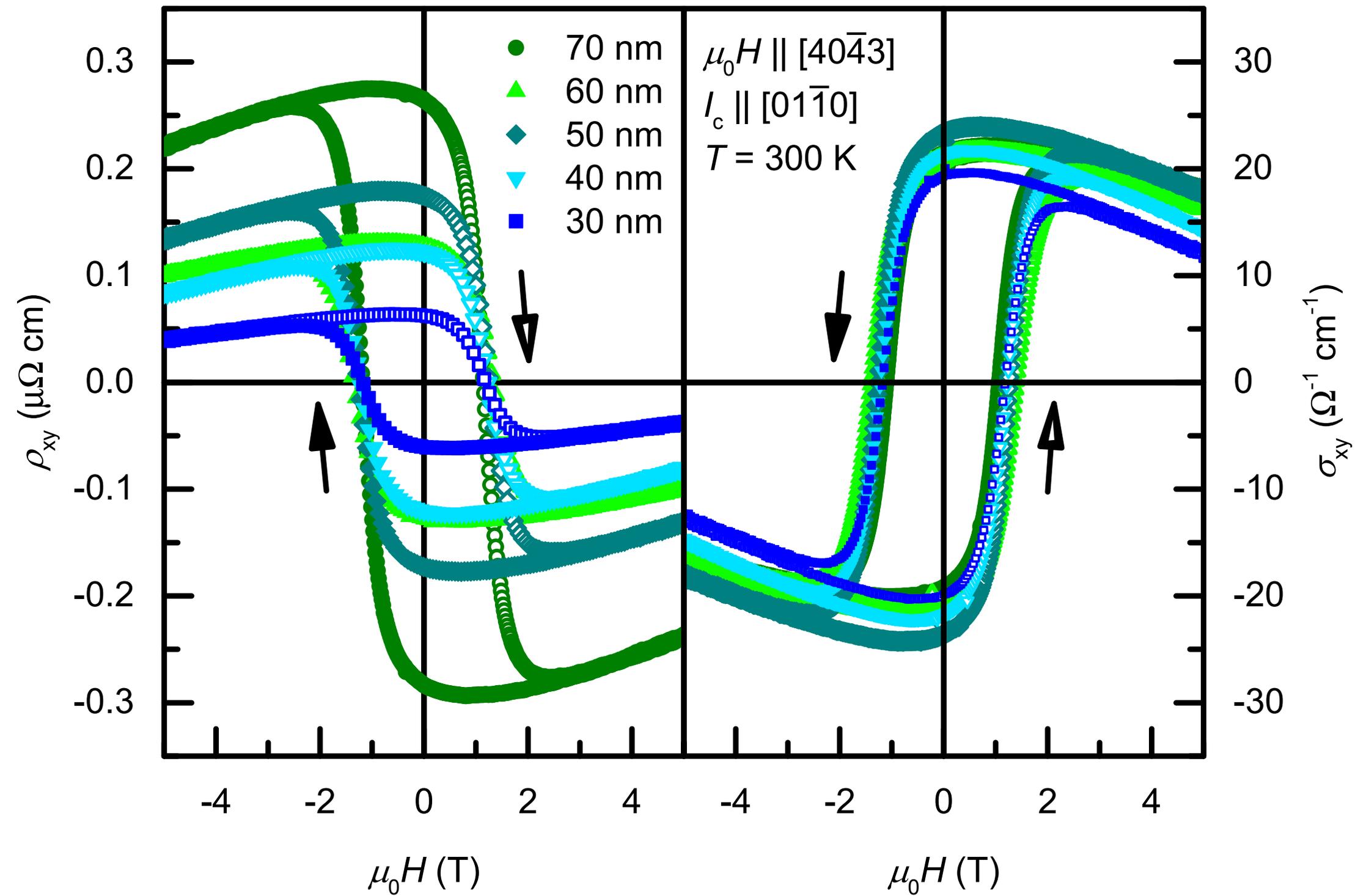

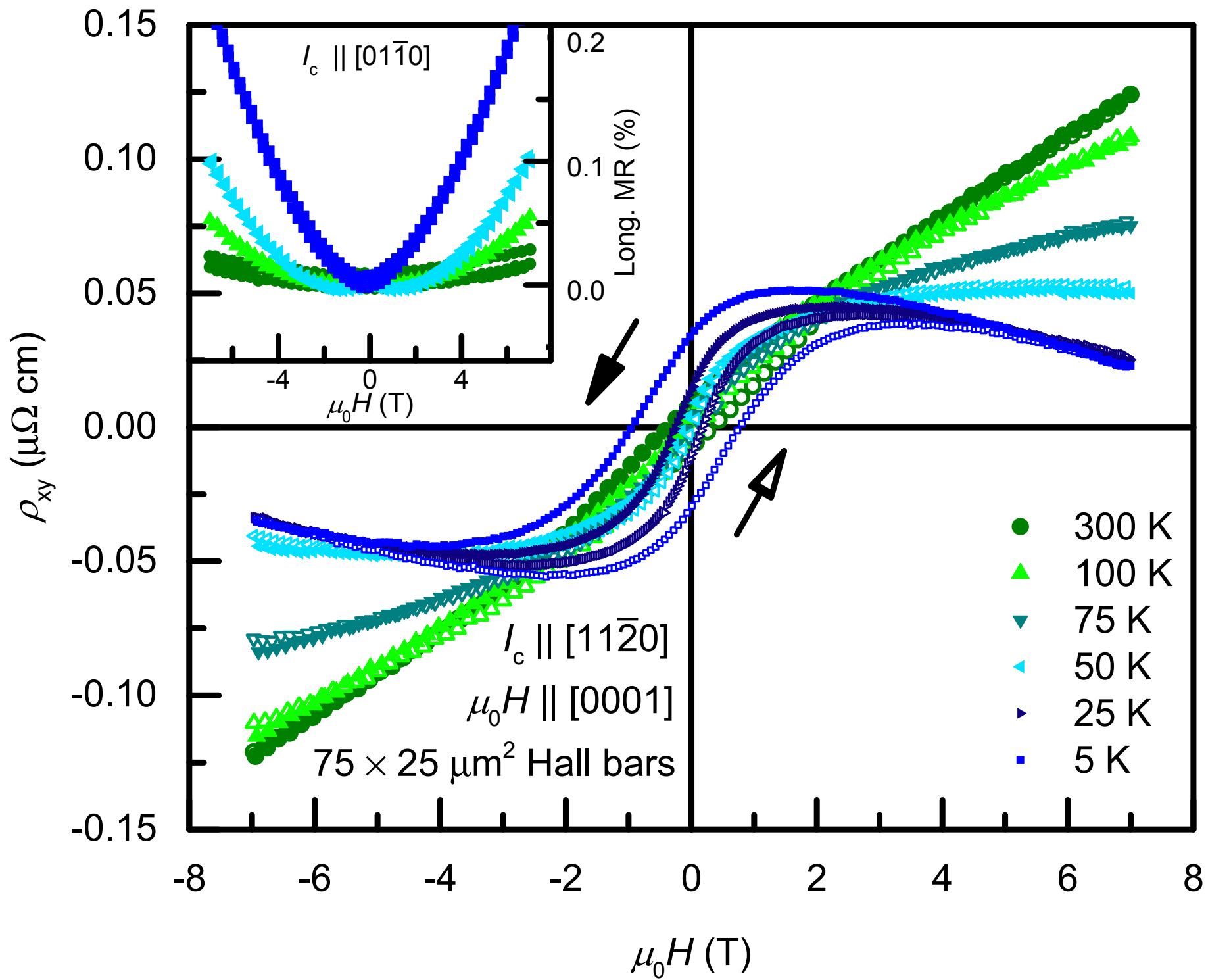

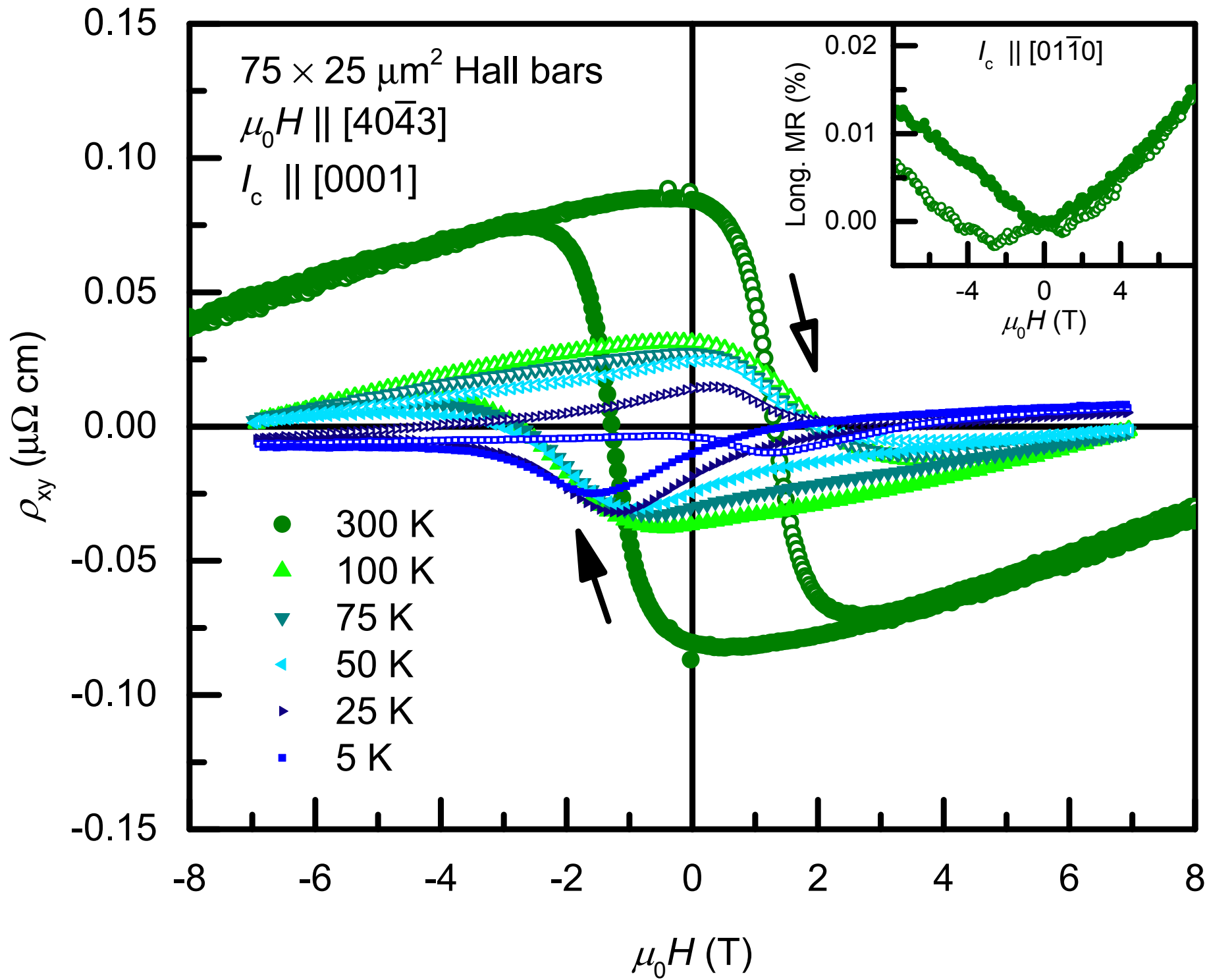

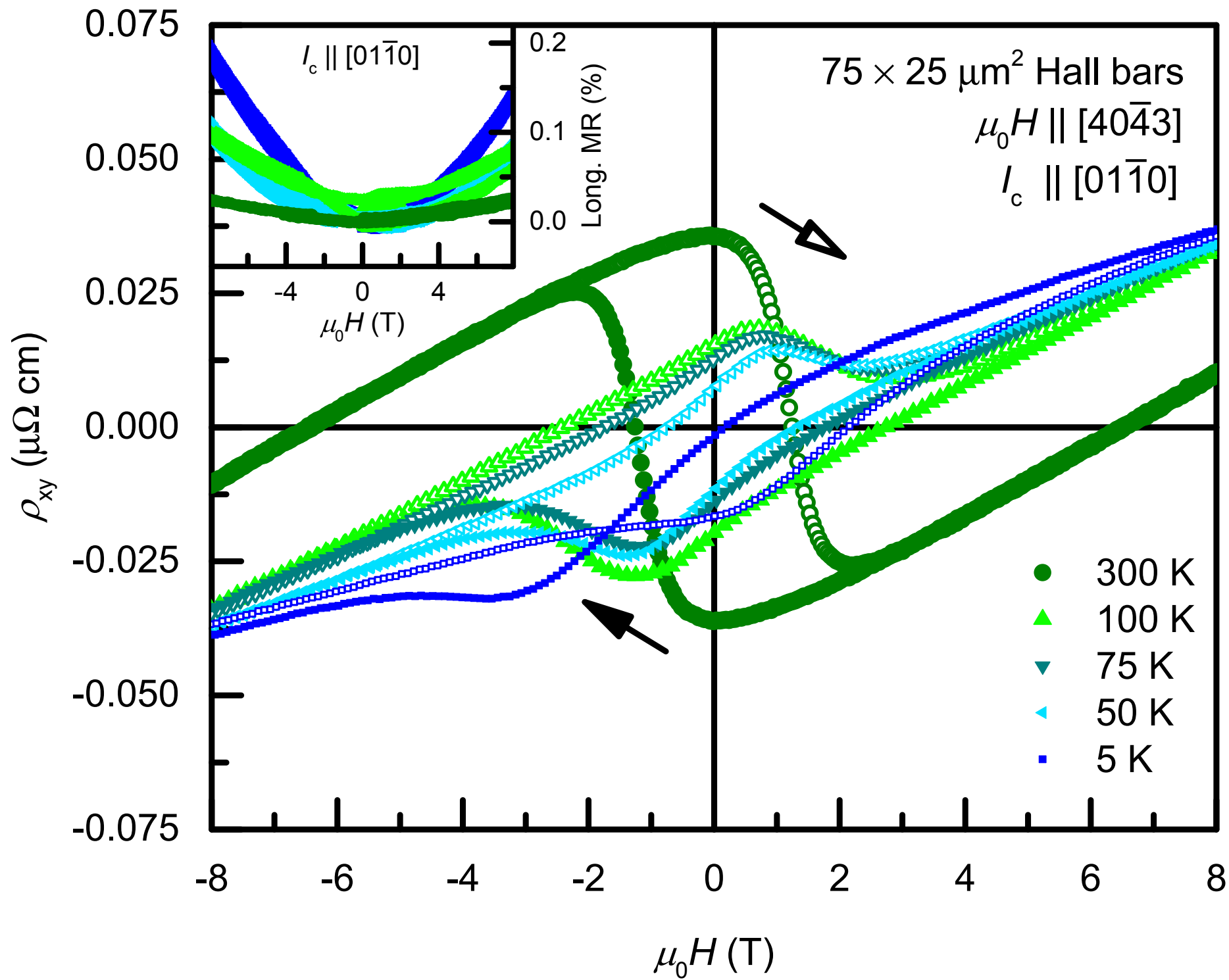

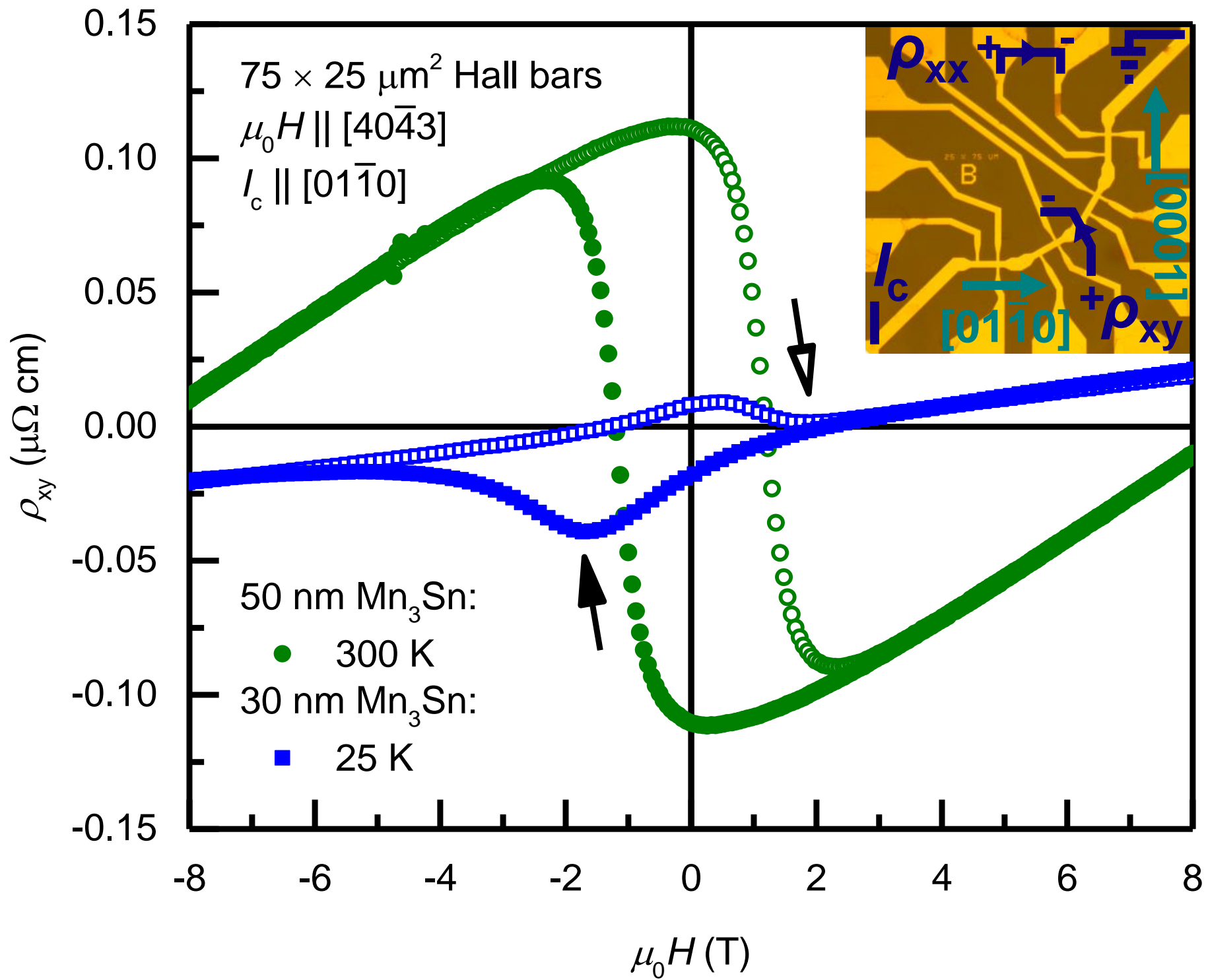

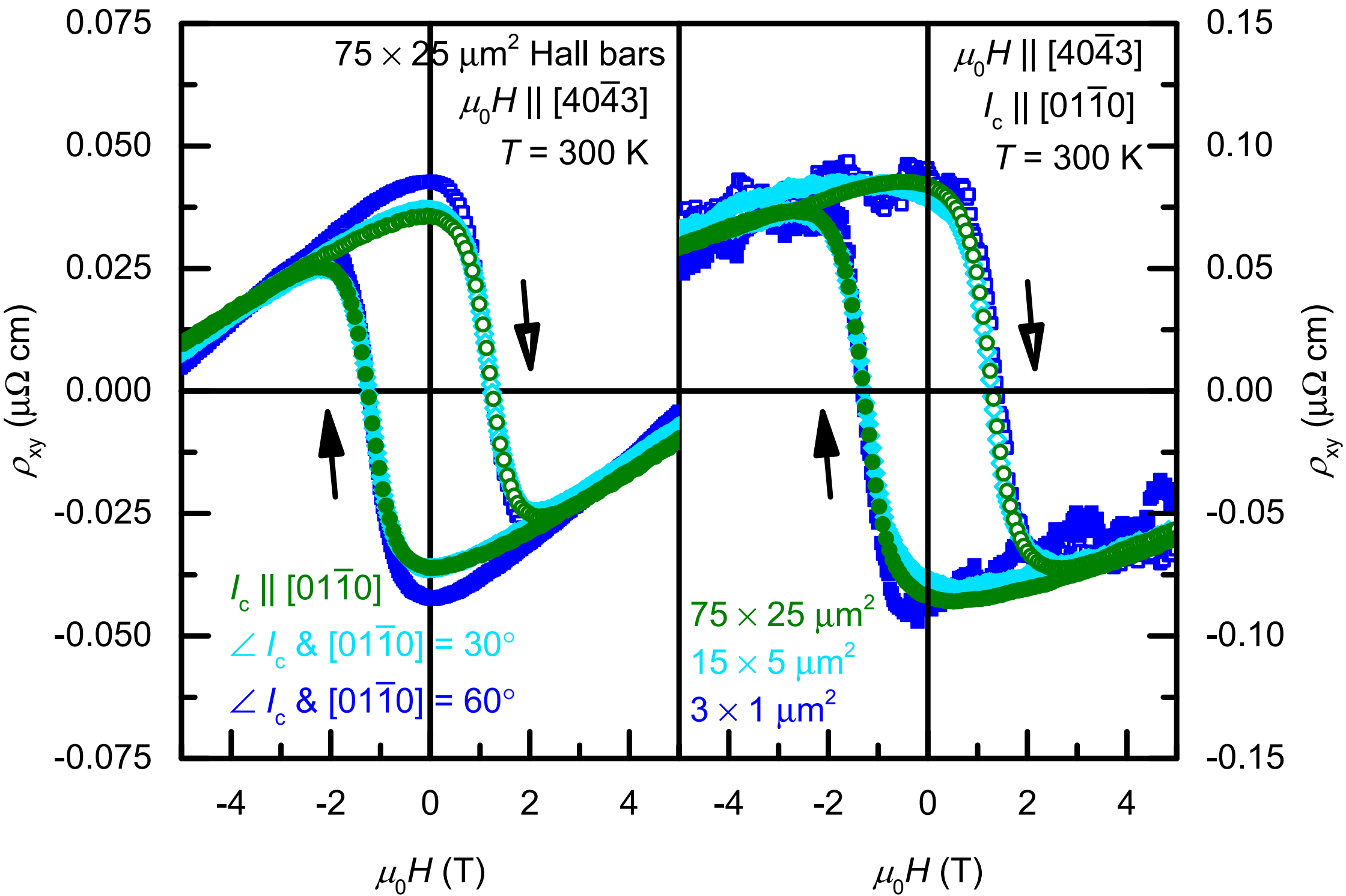